\documentclass[]{aa}  
\usepackage{natbib}
\bibpunct{(}{)}{;}{a}{}{,} 
\usepackage{graphicx}
\usepackage{xcolor}
\usepackage{txfonts}
\usepackage[colorlinks, allcolors=blue]{hyperref}
\DeclareUnicodeCharacter{2212}{-}
\begin{document}

   \title{Temporal evolution of small-scale internetwork magnetic fields in the solar photosphere}

   \author{R. J. Campbell
          \inst{1}
          \and
          M. Mathioudakis\inst{1}
          \and
          M. Collados\inst{2,3}
          \and
          P. H. Keys \inst{1}
          \and
          A. Asensio Ramos\inst{2,3}
          \and
          C. J. Nelson \inst{1}
          \and
          D. Kuridze\inst{4,5}
          \and
          A. Reid \inst{1}
          }

   \institute{Astrophysics Research Centre, Queen's University of Belfast, Northern Ireland, BT7 1NN, UK \\ email: rcampbell55@qub.ac.uk \vspace{-0.3cm}\\
    \and
        Instituto de Astrof\'isica de Canarias, V\'ia L\'actea s/n, E-38205 La Laguna, Tenerife, Spain\vspace{-0.3cm}\\
    \and
        Dept. Astrof\'isica, Universidad de La Laguna, E-38205, La Laguna, Tenerife, Spain \vspace{-0.3cm}\\
    \and
        Department of Physics, Aberystwyth University, Ceredigion, Cymru, SY23 3BZ, UK \vspace{-0.3cm}\\
    \and
        Abastumani Astrophysical Observatory, Mount Kanobili, 0301 Abastumani, Georgia \vspace{-0.3cm}
     \\
      }

   \date{Received 30 November 2020 / accepted 31 January 2021}
 
  \abstract
   {While the longitudinal field that dominates in photospheric network regions has been studied extensively, small-scale transverse fields have recently been found to be ubiquitous in the quiet internetwork photosphere and this merits further study. Furthermore, few observations have been able to capture how this field evolves.}
   {We aim to statistically characterize the magnetic vector in a quiet Sun internetwork region and observe the temporal evolution of specific small-scale magnetic features.}
   {We present two high spatio-temporal resolution observations that reveal the dynamics of two disk-centre internetwork regions taken by the new GRIS-IFU (GREGOR Infrared Spectrograph Integral Field Unit) with the highly magnetically sensitive photospheric Fe $\textsc{I}$ line pair at $15648.52$ $\textrm{\AA}$ and $15652.87$ $\textrm{\AA}$. 
We record the full Stokes vector and apply inversions with the Stokes inversions based on response functions (SIR) code to retrieve the parameters characterizing the atmosphere. We consider two inversion schemes: scheme 1 (S1), where a magnetic atmosphere is embedded in a field free medium, and scheme 2 (S2), with two magnetic models and a fixed $30\%$ stray light component.}
   {
   The magnetic properties produced from S1 inversions returned a median magnetic field strength of 200 and 240 G for the two datasets, respectively. We consider the median transverse (horizontal) component, among pixels with Stokes $Q$ or $U$, and the median unsigned longitudinal (vertical) component, among pixels with Stokes $V$, above a noise threshold. We determined the former to be $263$ G and $267$ G, and the latter to be $131$ G and $145$ G, for the two datasets, respectively. Finally, we present three regions of interest (ROIs), tracking the dynamics of small-scale magnetic features. We apply S1 and S2 inversions to specific profiles of interest and find that the latter produces better approximations when there is evidence of mixed polarities. We find patches of linear polarization with magnetic flux density of the order of $130−150$ G and find that linear polarization appears preferentially at granule-intergranular lane (IGL) boundaries. The weak magnetic field appears to be organized in terms of complex `loop-like' structures, with transverse fields often flanked by opposite polarity longitudinal fields.}
   {}

   \maketitle
\section{Introduction}

\begin{table*}
\caption{Atomic parameters of the five infrared Fe $\textsc{I}$ absorption lines in the observed spectral window. The laboratory rest wavelength is denoted by $\lambda_0$, while $\chi_l$ and log$(gf)$ denote the excitation potential of the lower level and the
logarithm of the multiplicity of the level upon the oscillator strength, respectively. The $\sigma_b$ and $\alpha_b$ are collisional broadening parameters
from the quantum mechanical theory of Anstee, Barklem, and O’Mara, with $\sigma_b$ given in units of the Bohr radius, $a_0$. $^{\mathrm{(a)}}$ Values taken from \cite{nave1994}. $^{\mathrm{(b)}}$ Values taken from \cite{borrero2003}. $^{\mathrm{(c)}}$ Values taken from \cite{bloomfield2007}. $^{\mathrm{(d)}}$ Values taken from \cite{ivan2019}. }       
\centering                          
\begin{tabular}{c c c c c c c c}        
\hline               
  \multicolumn{1}{|c|}{Ion} &$\lambda_0$ [$\AA$]& $g_{\mathrm{eff}}$ & $\chi_l$ [eV] & log($gf$) & $\sigma_b$ ($a_0$) & $\alpha_b$ & \multicolumn{1}{c|}{Electronic Configuration}  \\   
\hline 
 \multicolumn{1}{|c|}{Fe $\textsc{I}$} & 15645.02$^{\mathrm{(a)}}$ & 2.1$^{\mathrm{(d)}}$& 6.31$^{\mathrm{(a)}}$ & -0.65$^{\mathrm{(d)}}$ & 1035$^{\mathrm{(d)}}$ & 0.291$^{\mathrm{(d)}}$ & \multicolumn{1}{c|}{$^7$P$_2$ - $^7$P$_2$$^{\mathrm{(a)}}$} \\
 
 \multicolumn{1}{|c|}{Fe $\textsc{I}$} & 15648.52$^{\mathrm{(a)}}$ & 3.0$^{\mathrm{(b)}}$& 5.43$^{\mathrm{(a)}}$ & -0.675$^{\mathrm{(b)}}$ & 977$^{\mathrm{(b)}}$ & 0.229$^{\mathrm{(b)}}$ & \multicolumn{1}{c|}{$^7$D$_1$ - $^7$D$_1^{\mathrm{(a)}}$}\\
 
 \multicolumn{1}{|c|}{Fe $\textsc{I}$} & 15652.87$^{\mathrm{(a)}}$ & 1.5$^{\mathrm{(b)}}$& 6.25$^{\mathrm{(a)}}$ & -0.043$^{\mathrm{(b)}}$ & 1445$^{\mathrm{(b)}}$ & 0.33$^{\mathrm{(b)}}$ & \multicolumn{1}{c|}{$^{7}$D$_{5}$ - $^{6}$D$_{4.5}$ $4f[3.5]^{0}$$^{\mathrm{(a)}}$} \\
 
 \multicolumn{1}{|c|}{Fe $\textsc{I}$} & 15662.02$^{\mathrm{(a)}}$ & 1.5$^{\mathrm{(c)}}$& 5.83$^{\mathrm{(a)}}$ & 0.19$^{\mathrm{(c)}}$ & 1200$^{\mathrm{(c)}}$ & 0.239$^{\mathrm{(c)}}$ & \multicolumn{1}{c|}{$^5$F$_5$ - $^5$F$_4^{\mathrm{(a)}}$} \\
 
 \multicolumn{1}{|c|}{Fe $\textsc{I}$} & 15665.25$^{\mathrm{(a)}}$ & 0.8$^{\mathrm{(c)}}$& 5.98$^{\mathrm{(a)}}$ & -0.42$^{\mathrm{(c)}}$ & 1283$^{\mathrm{(c)}}$ & 0.234$^{\mathrm{(c)}}$ & \multicolumn{1}{c|}{$^5$F$_1$ - $^5$D$_1^{\mathrm{(a)}}$} \\
 
\hline 
\label{table:atomicdata}
\end{tabular}
\end{table*}

Granulation is the dominant pattern observed in the quiet solar photosphere, generated by convective cells rising from the convection zone. The photospheric `magnetic carpet' is continuously replenished, with flux in the photospheric network and internetwork regions balanced by the processes of flux emergence, fragmentation, coalescence and cancellation \citep{Gosic2014}. These processes have a role in determining the heating and dynamics of the upper atmosphere \citep{SCHRIJ1997}. While the longitudinal field that dominates network regions has been studied extensively, small-scale horizontal fields  ubiquitous in the internetwork photosphere merit more thorough investigation. Ultimately, developing an understanding of and ability to decipher the turbulent, mixed-polarity or `hidden' photospheric magnetic field, which is revealed by the Hanle effect to be present on small scales, could contribute to solving key problems in solar physics, such as heating of the upper atmosphere \citep{trujillo2004}. This magnetic field is hidden to the Zeeman effect at low spatial resolutions, and therefore requires very high-resolution observations \citep{BellotRubio2019}. As the polarization signals produced by weak fields have such low amplitudes, the Zeeman sensitivity of the employed spectral lines and signal-to-noise (S/N) of the observations is also of critical importance. The ratio of the Zeeman splitting, $\lambda_B$, to the Doppler width, $\lambda_d$, of a given spectral line, $\lambda_B/\lambda_d$, provides a measure of its sensitivity to different field strengths. The $\lambda_{B}$, may be expressed as
\begin{equation}\label{eqn:VtoB}
    \lambda_B = \frac{e}{4\pi mc}g_{\text{eff}}B\lambda_{0}^2 \approx 4.6686\times 10^{-13}g_{\text{eff}}B\lambda_{0}^2,
\end{equation}
where $e$ and $m$ are the charge and mass of an electron, respectively, $c$ is the speed of light, $g_{\text{eff}}$ is the effective Land\'e g-factor of the spectral line, $B$ is the magnetic field strength and $\lambda_{0}$ is the rest wavelength of the spectral line, when $B$ is expressed in G and $\lambda_B$ and $\lambda_0$ are expressed in $\AA$. Evidently, therefore, maximum Zeeman sensitivity is achieved with a weak line with large $g_{\text{eff}}$ in the infrared \citep{solanki1992}.

The Advanced Stokes Polarimeter (ASP) at the Dunn Solar Telescope (DST) provided some of the first spectropolarimetric observations of horizontal fields in the quiet solar photosphere \citep{lites1996}, revealing weak (typically $0.1-0.2\%$ of the continuum intensity), small (typically $<1-2''$), and transient (typically lasting $5$ mins) linear polarization signatures associated with magnetic inclinations parallel to the surface. A major advancement was achieved with the Spectro-Polarimeter (SP) instrument on board the space-based Hinode Solar Optical Telescope (SOT), revealing an internetwork region seemingly dominated by horizontal fields \citep{lites2008}. Using Zeeman-induced polarization diagnostics of the Fe $\textsc{I}$ $6301.5$ $\textrm{\AA}$ and $6302.5$ $\textrm{\AA}$ line pair in the visible ($g_{\text{eff}}$  of $1.67$ and $2.5$, respectively), the average apparent horizontal magnetic flux density was determined to be five times larger than the average apparent longitudinal flux density. The detected horizontal fields exhibited field strengths of $100-200$ G and small magnetic filling factors, $\alpha$. The vertical fields appeared concentrated in inter-granular lanes (IGLs), while the transverse fields were found at the edges of bright granules. Additionally, there were large regions, termed `voids', devoid of significant magnetic flux density. While the scans were conducted over a large field of view (FOV), this did not allow for the evolution of the magnetic field to be observed temporally. Therefore, observations were conducted in `sit-and-stare' or `deep' mode, allowing the authors to conclude the voids were in fact only regions of relatively smaller apparent flux density. \cite{lites2017} have since determined by examining only the linear polarization signals and their variance across the solar disk that the internetwork photosphere is dominated by weak horizontal fields.

Inversions have been applied to Hinode/SP data based on Milne-Eddington (ME) models, with some controversy. \cite{borrero2011} determined that an improper choice of selection criteria can result in the retrieval of a probability distribution for magnetic field inclinations, $\gamma$, defined as the angle between the magnetic vector and solar normal (and the observer's LOS at disk-centre), with a peak at $90^\circ$ due to noise contamination. \cite{ramos2009} employed a complete Bayesian analysis for ME atmospheres, concluding that the data contained enough information to constrain $\alpha$ and place upper limits on $B$. While $\gamma$ was constrained for pixels with signal above $4.5\sigma$, it could not be tightly constrained for pixels below this threshold, instead only constraining $\gamma$ in terms of the number of pixels with inclined ($60^\circ < \gamma < 120^\circ$) and vertical ($\gamma < 60^\circ$ or $\gamma > 120^\circ$) fields to resemble a quasi-isotopic distribution. The vast majority of pixels were not found to contain enough information to constrain the azimuth, $\phi$ (this requires both Stokes $Q$ and $U$ above the noise threshold).

\begin{figure*}
\centering
\includegraphics[width=\textwidth]{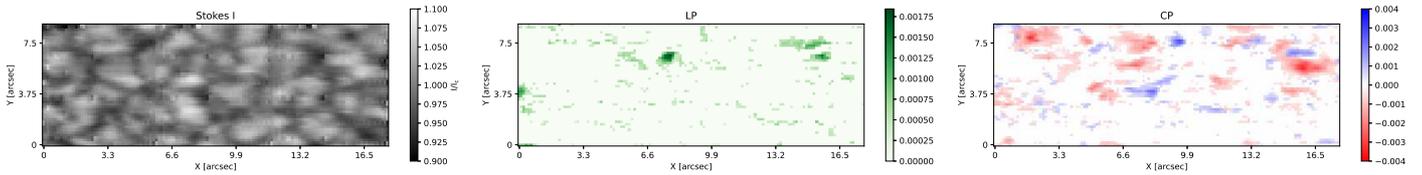}
\caption{Maps of Stokes $I$ (\textit{left}),
wavelength-integrated linear polarization (LP, \textit{middle}) and wavelength-integrated circular polarization (CP, \textit{right}) from a GRIS-IFU scan of a quiet Sun region taken at 07:55:24 UT on the $5$ May $2019$. Stokes $I$ is normalized by the average continuum. LP and CP are defined in the text. The data shown are reconstructed by the process described in section \ref{section:reconstruction}, and any Stokes parameter with maximum amplitude $<\sigma_t$ as measured in the continuum of the original level 1 data has been set to zero before computation of LP and CP. The CP map saturates at $\pm 0.004$ $I_\mathrm{c}$.}
          \label{fig:map}%
\end{figure*}

The exceptional Zeeman sensitivity of the Fe $\textsc{I}$ line pair at $15648.52$ $\textrm{\AA}$ and $15652.87$ $\textrm{\AA}$ - due to their relatively high $g_{\text{eff}}$ of $3$ and $1.5$, respectively, and near infrared wavelength - makes these lines an ideal Zeeman diagnostic for studying weak internetwork fields. \cite{khomenko2003} observed in the near infrared two very quiet regions at disk-centre, examining the statistical properties of these internetwork regions. The authors found that most of the observed fields were characteristically weak with relatively few kilo-Gauss features. Almost $30\%$ of the selected Stokes $V$ profiles were found to exhibit irregular shapes with multiple lobes and polarities: a characteristic signature of mixed line-of-sight (LOS) velocities and unresolved magnetic structures within the resolution element. \cite{almeida2003} were the first to simultaneously observe an internetwork region with the aforementioned Fe $\textsc{I}$ visible and near infrared line pairs. The visible lines were found to trace kilo-Gauss fields while the near infrared lines traced weaker fields. \cite{marian2008} also simultaneously observed an internetwork region with the same line pairs, and, conversely, found that the polarity of Stokes $V$ in both maps were aligned. Importantly, the near infrared lines were found to be more suitable for detection of horizontal fields due to their increased ability to measure linear polarization. \cite{khomenko2007} investigated the diagnostic capability of these two line pairs, finding that the differences may be attributed to a large difference in formation heights.

Disk-centre quiet sun imaging obtained by the Fabry-P\'erot Imaging Magnetograph Experiment (IMaX) \citep{IMaX2011} on board the balloon-borne SUNRISE observatory \citep{Sunrise2010} revealed magnetic structures at very high-resolution ($100$ km) and cadence ($33$ s) in the magnetically sensitive $5250.2$ $\textrm{\AA}$ line. \cite{dan2010} used an automated detection method to locate over $4000$ features with significant linear polarization, appearing preferentially at granule-IGL boundaries and occurring both in up-flows and down-flows. \cite{IMaX2012} identified nearly 500 small-scale magnetic loops in these data. The observations revealed large voids without significant magnetic flux. \cite{IMaX2013} extrapolated these photospheric measurements into the upper atmosphere and determined that magnetic reconnection likely cannot solely account for heating the chromosphere and corona in the quiet Sun. \cite{IMaX_marian_2011} determined that $B$ and the area of longitudinal magnetic flux patches oscillated in anti-phase in a manner consistent with granular motions. More recent analysis has focused on the linear polarization features (LPFs), with \cite{IMaX2018} determining them to be short-lived ($30-300$ s), small structures with weak fields observed equally in up-flow and down-flow regions. \cite{solanki2019} found cancellation sites in the quiet Sun surrounding a young active region. Linear polarization (i.e. horizontal fields) was found along the polarity inversion line (PIL) separating opposite polarity circular polarization (i.e. vertical fields). Detection of such potential cancellation sites in the internetwork photosphere remains a key goal \citep{anjaliIFU}.

%
\begin{table*}[t]
\caption{Time-averaged percentage of linear (LP) and circular (CP) polarization profiles
above given $\sigma-$thresholds for level 1 GRIS-IFU/GREGOR, GRIS/GREGOR and IMaX/SUNRISE data. The percentages are calculated relative to the full FOV. The $1\sigma$ noise level is determined by the calculation of the standard deviation in the relevant Stokes vector at continuum wavelength(s).}       
\label{table:1}      
\centering                          
\begin{tabular}{|c| c | c | c | c | c c c c|}        
\hline               
 Instrument & Date (yyyy-mm-dd) & Start time [UT] & Mean Q 1$\sigma$ value & $\sigma-$level & $\%$ LP & $\%$ CP & $\%$ LP + CP & $\%$ NP \\   
\hline      
    GRIS-IFU & 2019-05-05 &07:29:03 & $7.9\times10^{-4}$I$_{\textrm{c}}$ &  3.0 & 52.7 & 61.9 & 35.8 & 21.2  \\ 
    && & & 4.0 & 11.5 & 30.4 & 5.4 & 63.6 \\ 
    && & & 4.5 & 5.9 & 23.6 & 2.5 & 73.0 \\ 
\hline      
   GRIS-IFU & 2019-05-06 & 07:33:27& $8.1\times10^{-4}$ I$_{\textrm{c}}$ & 3.0 & 51.3 & 58.8 & 33.5 & 23.4  \\    
    && & & 4.0 & 12.9 & 29.3 & 6.0 & 63.8 \\ 
    && & & 4.5 & 6.6 & 21.3 & 2.7 & 74.8 \\ 
\hline                                   
   GRIS & 2015-09-17 & 08:26:50  & $4.6\times10^{-4}$I$_{\textrm{c}}$ & 3.0 & 42.3 & 50.1 & 23.0 & 30.7   \\
   && & & 4.5 & 8.3 & 26.8 & 3.2 & 68.1 \\ 

\hline
   IMaX & 2009-06-09 & 00:35:49 & $8.4\times10^{-4}$I$_{\textrm{c}}$ & 3.0 & 2.3 & 12.0 & 0.7 & 86.4  \\     

\hline
   IMaX & 2009-06-09 & 01:30:40 & $8.4\times10^{-4}$I$_{\textrm{c}}$ & 3.0 & 2.5 & 11.2 & 0.7 & 87.0  \\     

\hline  

\end{tabular}
\end{table*}

\cite{lagg2016} used the Fe $\textsc{I}$ $15648.52/15652.87 \textrm{\AA}$ lines with the GREGOR Infrared Spectrograph (GRIS)  to observe an internetwork region at disk-centre with very high spatial resolution. Following spatial and spectral binning it was revealed that $9.2\%$ of FOV showed linear polarimetric signals above a $5\sigma$ level. \cite{marian} inverted these data using two inversion schemes; first, a magnetic model embedded with a field-free model, and second, two magnetic model atmospheres with a stray light component. One half of their magnetized FOV was effectively modelled by the former scheme, while the other half was better modelled by the latter. However, the question remains unresolved as to whether this substructure is generated by gradients along the LOS or across the solar surface. \cite{kiess} inverted these data, focusing Stokes $V$ profiles with more than two lobes, indicative of the presence of opposite polarities co-existing either vertically (along the LOS) or horizontally within the resolution element.
   
This study will utilize Zeeman diagnostics to investigate the temporal evolution of weak, small-scale magnetic fields from the linear and circular polarization signals observed in the internetwork. This requires simultaneous achievement of adequate spatial resolution, polarimetric sensitivity, spectral resolution, and high cadence imaging that has previously been difficult to achieve. The new GRIS-IFU (GREGOR Infrared Spectrograph Integral Field Unit) instrument mounted at the $1.5$ m GREGOR telescope provides the ideal instrument for this purpose, with a relative compromise made on the FOV. These observations are the highest spatial resolution near infrared time-series imaging available to date. In section \ref{section:observations} we discuss the observational data and their reduction. In section \ref{section:results1} we analyse the results in terms of polarization amplitudes (section \ref{section:polarization}) and outline the strategy employed for noise reduction and correction of instrumental imperfections (section \ref{section:reconstruction}). In section \ref{section:inversions} we apply inversions to the data and discuss the results both in terms of overall statistics (section \ref{section:statisticalanalysis}) and select regions of interest (ROIs, section \ref{section:regionsofinterest}). In section \ref{section:conclusions} the results are discussed and the implications of the work are summarized.

\section{Observations}\label{section:observations}
On the $5$ and $6$ May $2019$ observations of two very quiet internetwork regions, very close to disk-centre were taken (recorded Helio-projective co-ordinates (HPC) were [x,y] $= [14, 49]''$, $\mu=\cos\theta=0.99$, and [x,y] $= [2, 5]''$, $\mu=1.00$, respectively, where $\theta$ is the heliocentric angle) with the GRIS-IFU (Dominguez-Tagle et al., in prep.) mounted at GREGOR \citep{Schmidt2012}. Co-ordinates were selected using Helioseismic and Magnetic Imager (HMI) magnetograms \citep{sdohmi}, however there is a large uncertainty in the pointing information at GREGOR and therefore we could not rule out the presence of network elements. The GRIS-IFU was operated in double sampling mode. An exposure time of $30$ ms per polarimetric state was chosen, with 10 accumulations, resulting in a cadence of $64$ seconds between frames. Scans were acquired from 07:29 to 08:34 UT and 07:33 to 08:33 UT on the $5$ and $6$ of May, respectively. The standard GRIS reduction pipeline \citep{gregor2012} was employed to the data for the purposes of dark current removal, flat fielding, polarimetric calibration and cross-talk removal (Stokes $I$ to Stokes $Q$, $U$, and $V$). It is known that residual cross-talk between the polarized Stokes vectors of a few percent could remain and the atmosphere may introduce additional seeing-induced cross-talk, but the latter is very unlikely to be larger than the former.

The GRIS-IFU was operated in the near infrared, observing a $40$ $\textrm{\AA}$ spectral window. Table \ref{table:atomicdata} shows atomic data for the five Fe $\textsc{I}$ lines in the observed spectral window. This spectral range contains several absorption lines, including the Fe $\textsc{I}$ line pair at $15648.52$ $\textrm{\AA}$ and $15652.87$ $\textrm{\AA}$. We degraded the spectral resolution of a Fourier Transform Spectrometer (FTS) atlas \citep{atlas} until it matched an average continuum-normalized Stokes $I$ quiet Sun profile and determined the spectral dispersion to be $39.58$ m$\AA$/pixel.

The GRIS-IFU builds up an image raster in a mosaic pattern, and for our observations a $3\times3$ mosaic was chosen as a compromise between cadence, integration time and FOV so that the effective FOV of each scan is $18''$ by $9''$ with a spatial sampling of $0.135''$/pixel by $0.188''$/pixel. Estimation of the effective spatial resolution of the observations from the power spectrum of the granulation at a continuum wavelength is not appropriate due to the small FOV. The peak root mean square continuum contrast of the granulation was measured as $2.8\%$ and $3.1\%$ for the $5$ and $6$ May scans, respectively. These values are slightly higher than the $2.3\%$ continuum contrast reported by \cite{lagg2016} for GRIS.

Figure \ref{fig:map} shows a sample frame from the $5$ May scan. A few pixels at the edges of the IFU tile were consistently over- or under-saturated. This impacts less than $4\%$ of the FOV. In Figure \ref{fig:map} we have interpolated to remove many of these pixels for visualization purposes only. We used a Sobel filter acting on Stokes $I$ at a continuum wavelength to detect and mask pixels containing artefacts. The Sobel operator convolves two $3\times3$ kernels with the time-averaged Stokes $I$ continuum image at each pixel to calculate approximations of the horizontal and vertical derivatives that can be combined to return the gradient magnitude at each point in the image. As the granulation evolves, and, thus, Stokes $I$ averages out to a large extent over the course of the time-series, by thresholding the gradient magnitude, we were able to identify those pixels with instrumental artefacts. These pixels are masked in any subsequent analysis reported in this study. Additionally, at the beginning of the $5$ May scan, the slit was visible in Stokes $I$ resulting in the loss of 13 frames of data. The polarization vectors seem to be unaffected, perhaps due to the fact that they are modulations of Stokes $I$, but the loss of Stokes $I$ means that these frames cannot be inverted. We identify 22 frames in each scan, from 07:45:46 to 08:08:16 UT on the $5$ May and from 07:33:27 to 07:55:59 UT on the $6$ May that have above average or good seeing conditions, quantified by the fried parameter, $r_0$, and are free from these artefacts and as such they are suitable for analysis. At GREGOR, an $r_0$ of $6$ cm or higher is considered average to good while above $8$ cm is considered good to exceptional. We average the values across the $9$ tiles in each frame and assign the average value.

\section{Results and analysis}
\subsection{Noise-treatment and polarization analysis}\label{section:results1}

\subsubsection{Temporal quantification of polarization signals}\label{section:polarization}
We defined the wavelength-integrated, net linear (LP) and circular (CP) polarization as follows:
\begin{equation}
    \textrm{LP} = \frac{\int_{\lambda_b}^{\lambda_r} [Q^2(\lambda)+U^2(\lambda)]^{\frac{1}{2}}  d\lambda}{I_c \int_{\lambda_b}^{\lambda_r} d\lambda},
\end{equation}
    
\begin{equation}
    \textrm{CP} = \textrm{sgn}(V_{\textrm{b}}) \frac{\int_{\lambda_b}^{\lambda_r} |V(\lambda)|  d\lambda}{I_c \int_{\lambda_b}^{\lambda_r} d\lambda},
\end{equation}
where $\textrm{sgn}(V_{\textrm{b}})$ represents the sign of the blue Stokes $V$ lobe, $\lambda_r$ and $\lambda_b$ are the red and blue wavelength limits of integration, respectively, and $I_c$ is the continuum intensity. We followed the scheme of \cite{lagg2016} in quantifying the number of pixels exhibiting a polarization signal above the noise level, so that we could compare our results to other datasets. Firstly, for a given Stokes vector, to calculate the $1\sigma$ noise level, we calculated the standard deviation of that Stokes parameter in the continuum, under the assumption that the continuum is unpolarized. Next, the Stokes parameters were thresholded accordingly, so that if, for a given pixel, the Stokes $Q$ or $U$ signal possessed a maximum value across the spectral line greater than given multiples of $\sigma$, the pixel was considered to exhibit an LP signal, and similarly for Stokes $V$ and CP. If a given pixel did not meet the required threshold in Stokes $Q$, $U$, or $V$, it was said to have no polarization (NP). 

The time-averaged results of this quantification analysis are shown in table \ref{table:1}. For comparison, also reported in Table \ref{table:1} are the statistics calculated for two of the most relevant spectropolarimetric datasets available of disk-centre internetwork regions, namely a scan taken by GRIS/GREGOR using the same near infrared spectral region reported in this study \citep{lagg2016} and a dataset taken by the balloon-borne IMaX/SUNRISE experiment using the less magnetically sensitive $5250.2$ $\AA$ Fe $\textsc{I}$ line \citep{Sunrise2010}. Here we made use of the IMaX data before phase diversity reconstruction as while this would improve the effective spatial resolution, it would also have a significant detrimental impact on the noise level \citep{shahin2014}. As the effective spatial resolution of the level 1 IMaX dataset is nevertheless still higher resolution than the GRIS-IFU data, and the two datasets have similar noise levels, we therefore chose not to employ phase diversity reconstruction. While similar statistics for these datasets can be found reported in previous studies, each study applies a different treatment to the level 1 data - usually some combination of spatial or spectral re-binning, sometimes at the cost of respective resolution - in an attempt to reduce the noise, and therefore we present this analysis for each of these datasets without any binning to enable a valid comparison. Although the probability of photon noise producing a signal above $3\sigma$ for one wavelength point is $0.3\%$, the chance for this to occur across the spectral line, with many wavelength points, is much higher \citep{borrero2011}. For IMaX, with relatively few sampling points across the line, $3\sigma$ is likely sufficient and clearly selects real polarization features. However, for the GRIS and GRIS-IFU a more stringent $4-4.5\sigma$ threshold is required to select similar features. There is a reasonable similarity in the average statistics between both of the GRIS-IFU scans at $4.5\sigma$, which could be interpreted as evidence that the observed regions are both characteristic of the quiet Sun internetwork, or at least that similar targets were observed. Importantly, the $1\sigma$ noise level for the GRIS scan is lower than for the GRIS-IFU scans, which helps explain the small differences at $4.5\sigma$ between these instruments.

\subsubsection{Reconstruction of the Stokes profiles}\label{section:reconstruction}

\begin{figure}

\includegraphics[width=\columnwidth]{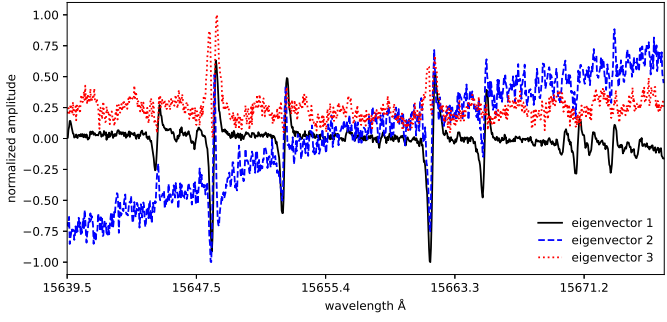}

\caption{First three eigenvectors in the base for PCA for Stokes $Q$, revealing the presence of intensity gradients and interference fringes. Each eigenvector is shown with a normalized amplitude for visualization purposes.}
          \label{fig:eigenvectors}%
\end{figure}

\begin{figure}
\includegraphics[width=\columnwidth]{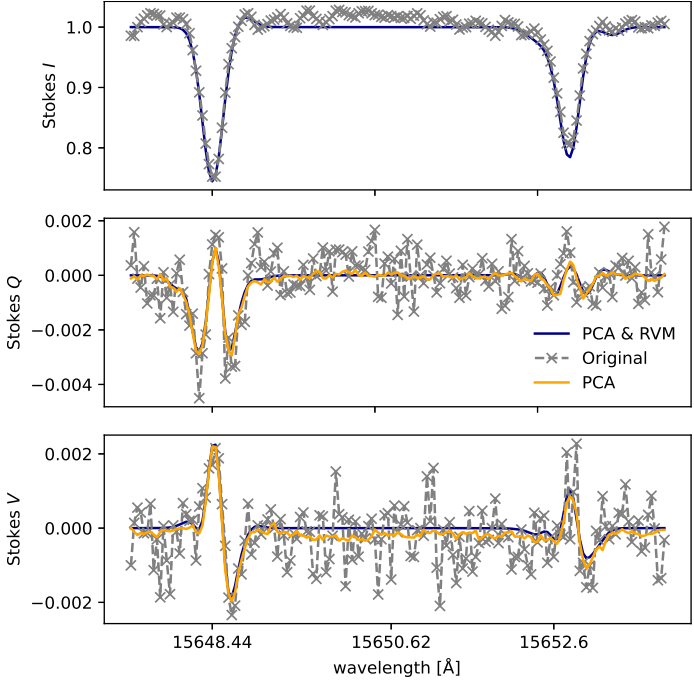}
\caption{Sample profile with Stokes $Q$ and $V$ signal after filtering by PCA with 15 retained eigenvectors and subsequent zero-shift correction and reconstruction by the RVM.}\label{fig:exampleprofilePCARVM}%
\end{figure}

Without precise determination of the effective spatial resolution, spatial and spectral binning can be undesirable as it may result in mixing of opposite polarity signals and thus modification of the Stokes vectors (typically a reduction in amplitude). The amplitudes of the polarization profiles are characteristically weak. The maximum amplitude of Stokes $V$ recorded in the $5$ May scan was 0.02 $I_\mathrm{c}$, while the maximum Stokes $Q$ and $U$ amplitude was lower at 0.009 $I_\mathrm{c}$ and 0.008 $I_\mathrm{c}$, respectively. For such weak profiles, noise can influence the results obtained to a large degree, especially when considering inversions. We therefore sought to reduce the noise using the same principle component analysis (PCA) \citep{rees2000, rees2004} strategy as has been employed in many recent studies (e.g. \citealt{khomenko2003}, \citealt{marian}). Assuming we have a set of observations, $S_\textrm{lj} = S_l (\lambda_j)$, with a given number of wavelengths, $j = 1, ..., N$, and a given number of pixels, $l = 1, ..., M$, for each Stokes vector, $S$, each profile is defined as a linear combination of a set of orthonormal eigenvectors, $e_i$, $i = 1, ..., n$,
\begin{equation}
    S (\lambda_j) = \sum_{i=1}^{n} c_i e_i (\lambda_j),
\end{equation}
where $c_i$ are the respective eigenvector coefficients. We directly obtained a set of eigenvectors and coefficients using a single value decomposition (SVD) solution \citep{press}, providing an orthogonal vector base in which the eigenvectors were ordered according to the decreasing non-negative amplitude of the corresponding singular values, reflecting each vector's relative importance of their contribution to the construction of each eigenprofile. When most of the singular values are small, $S (\lambda_j)$ can be well represented by only a few terms in the sum. Indeed, as it turns out, only a finite number of eigenvectors contain information necessary to reproduce the spectral line profiles, while the rest contain information necessary to reproduce the noise pattern of each profile. By truncating each profile after a given number of eigenvectors, we were therefore able to `de-noise' the data. If the number of retained eigenvectors is correctly chosen, this can be achieved with minimal loss of signal. The base was built from a frame that contained the fewest artefacts and the best seeing conditions. For each Stokes parameter, the spectra were ordered according to their maximum amplitude and the 1500 strongest profiles were chosen to form the base. After testing the application of PCA with a variable number of retained eigenvectors, and subtracting the original and reconstructed datasets to ensure the magnitude of the difference in LP and CP for those pixels with signal greater than $3\sigma$ was always of the order of or lower than the noise, we determined that the number of eigenvectors that should be retained for Stokes $Q$, $U$, and $V$ was 15.

\begin{figure}
   \centering
   \includegraphics[width=\columnwidth]{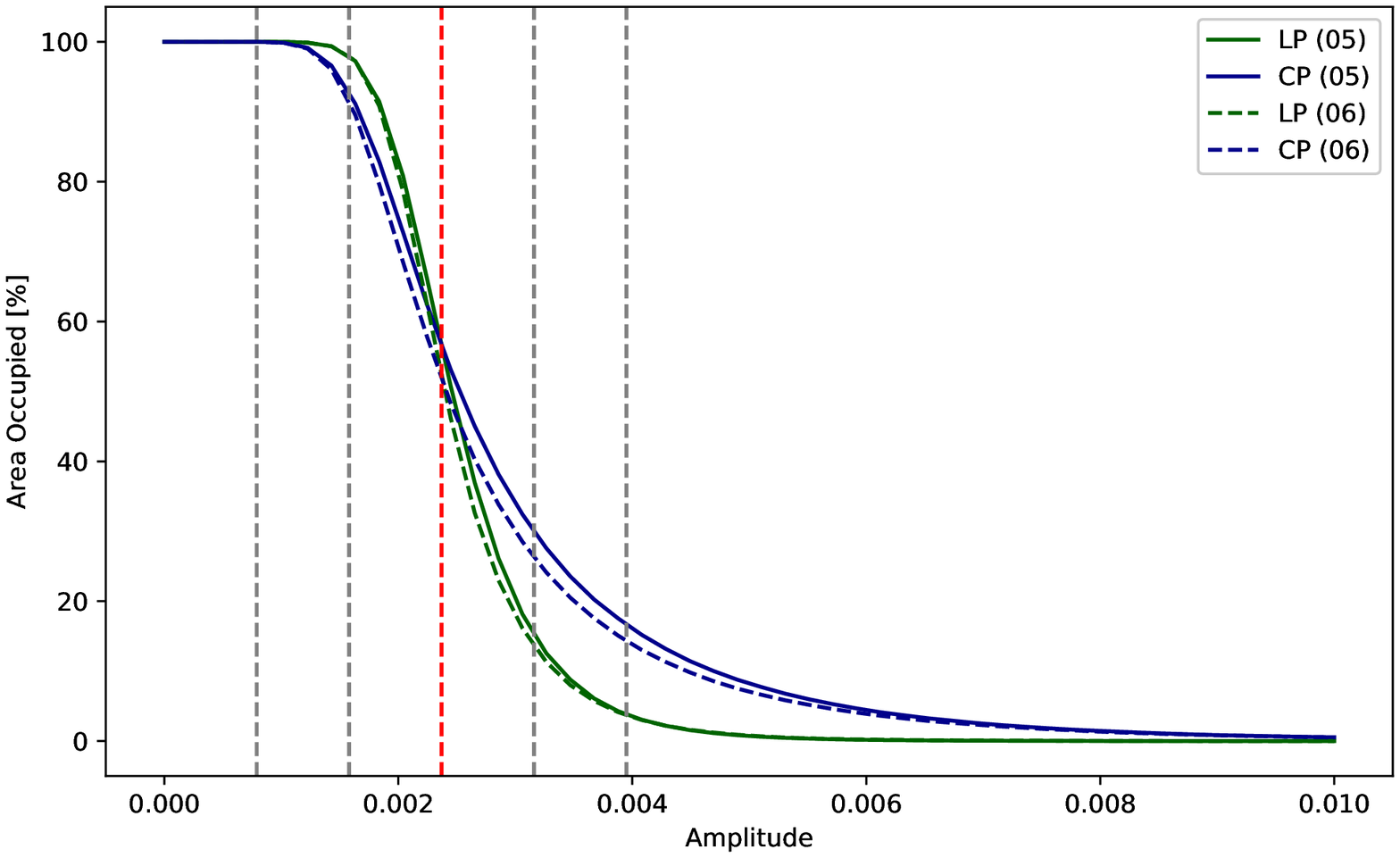}
   \includegraphics[width=\columnwidth]{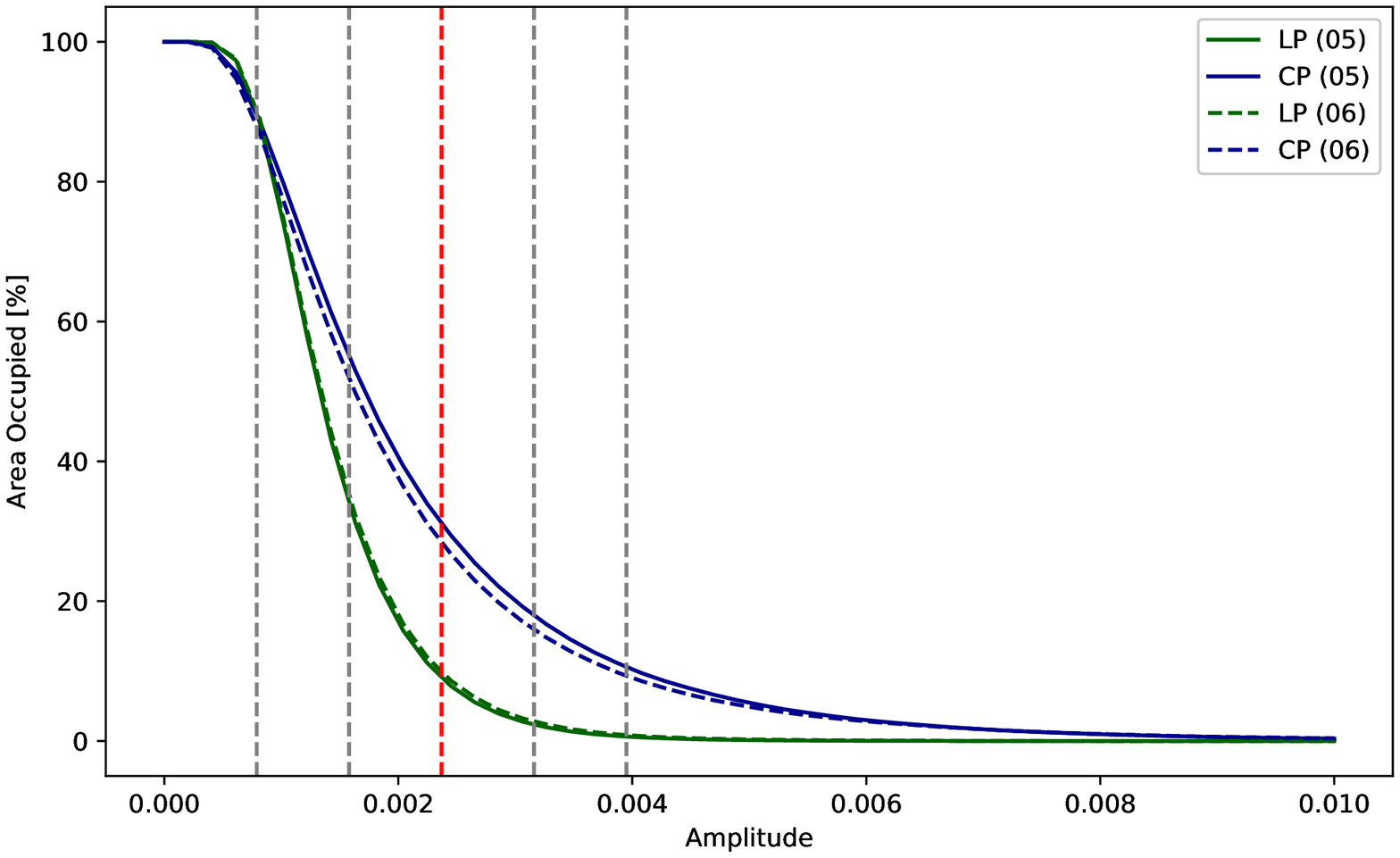}
   \includegraphics[width=\columnwidth]{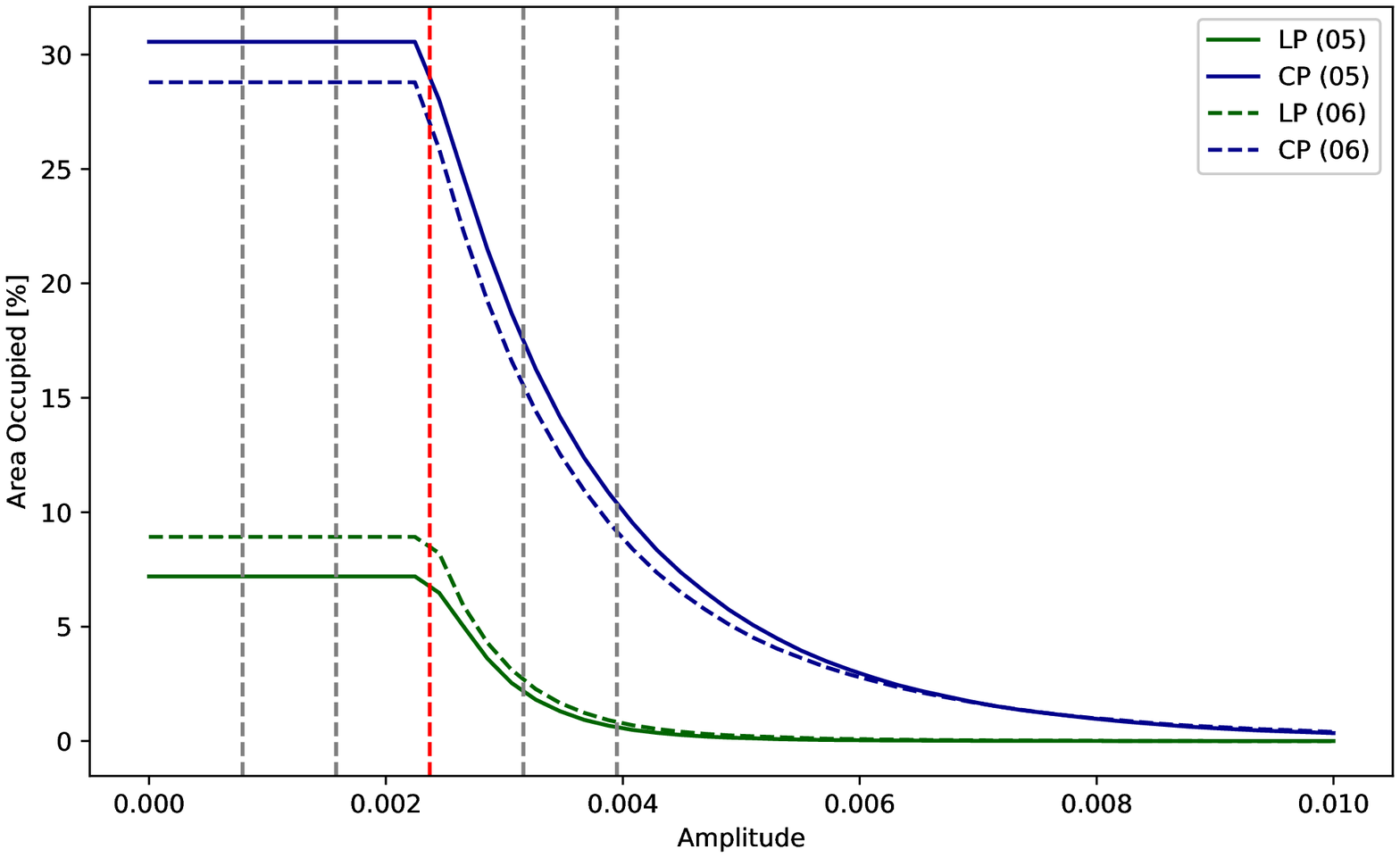}

   \caption{Total area occupied by pixels with maximum amplitude greater than given amplitudes for the scans on the $5$ May (\textit{solid lines}) and $6$ May (\textit{dashed lines}). A pixel has an LP above a given amplitude if its maximum Stokes $Q$ or $U$ value exceeds that amplitude across the $15648.52\AA$ line (\textit{green lines}) and likewise for CP and Stokes $V$ (\textit{blue lines}). \textit{Upper:} Original level 1 data. \textit{Middle:} Data after PCA filtering. \textit{Lower:} Data after PCA filtering, reconstruction by the RVM and with any Stokes vector with maximum amplitude $< \sigma_t$ (\textit{vertical, red, dashed line}) set to zero. The \textit{vertical} \textit{dashed} \textit{lines} indicate the $1, 2, 3, 4$ and $5\sigma$ (\textit{left} to \textit{right}) levels averaged across all frames as measured in the original data. The $\sigma$ level is averaged for all pixels in a given frame and is specific to a given Stokes vector.}
              \label{fig:statsPCARVM}%
    \end{figure}

The results of PCA revealed the presence of polarized and unpolarized interference fringes and intensity gradients that were previously not noticed (see Figure \ref{fig:eigenvectors}). In addition, many of the PCA-filtered polarized Stokes profiles were not centred at zero counts in the continuum. One approach to removing the former artefacts is to fit sinusoidal functions and low-order polynomials to the eigenvectors, remove them from the eigenvectors and then, crucially, re-orthogonalize the base, as in \cite{marian}. However, we decided to take a different approach that additionally produces apparently noiseless data. First, we corrected the wavelength-dependent zero-shift for the polarized Stokes vectors by determining the constant $\delta$,
\begin{equation}
    \delta =\frac{\Delta S(\lambda_j)}{I(\lambda_j)},
\end{equation}
where $\Delta S_j$ is the difference between the amplitude of the Stokes vector and zero, $I_j$ is Stokes $I$, each at a given wavelength ($|\overline{\delta}|=1.3\times10^{-4}, 1.0\times10^{-4}, 1.0\times10^{-4}$ for Stokes $Q$, $U$, and $V$ respectively in the $5$ May scan, averaged across all pixels). Second, we employed a sparse Bayesian regression model \citep{tipping}, known as a relevance vector machine (RVM), to provide an approximation to each Stokes profile using a linear function comprising a small number of fixed basis functions from a large dictionary of potential candidates,
\begin{equation}
    S(\lambda_j) = \sum_{l=1}^{M}\omega_l \phi_l (\lambda_j) + \epsilon,
\end{equation}
under a sparsity constraint, where only a few of the weights, $\omega_l$, to the functions, $\phi_l$ are non-zero, and with the addition of some error, $\epsilon$, under the assumption that the errors are modelled as zero-mean Gaussians. The noise level can be set in advance or estimated from the data. The RVM exploits properties of the marginal likelihood function to enable efficient sequential addition and deletion of candidate basis functions. The size of the basis set, or dictionary of functions, is variable. We introduced: sinusoidal functions with periods in a range centred on those measured from the interference fringes in the PCA eigenvectors, gaussians at wavelengths where we have spectral lines that we wish to retain, and with widths in a range centred on the full width at half maximums (FWHMs) of each spectral line, and low-order polynomials to fit and remove spectral gradients.
We then exploited the linearity of the functions to select and delete those that we wanted to remove, in particular the interference fringes and spectral gradients. Only the five Fe $\textsc{I}$ spectral lines listed in Table \ref{table:atomicdata} were retained.

Figure \ref{fig:exampleprofilePCARVM} shows the total collapse in noise achieved via application of this full reconstruction scheme for a sample profile with Stokes $Q$ and $V$ signal just above the noise level. Figure \ref{fig:statsPCARVM} shows the area occupied by pixels with linear and circular polarization signals above sequentially increasing amplitudes before and after reconstruction. A large number of randomized profiles were manually checked to confirm the effectiveness of the RVM on reconstructing the Stokes profiles. For all profiles above $3\sigma$ and $5\sigma$ in Stokes $Q$ or $U$ in the $5$ May scan, the mean percentage difference in LP across the FOV of all frames was $11.1\%$ and $7.3\%$, respectively. Given that the RVM is removing spectral gradients and interference fringes, a small difference is to be expected. However, evidently, the RVM performs worse for noisier profiles. We consider the magnitude of this difference to be acceptable because, as in the example shown in Figure \ref{fig:exampleprofilePCARVM}, the shape and separation of the Stokes profiles is remarkably well reconstructed. The Stokes $I$ continuum is flattened, which is essential as this level contains important thermodynamic information, except for a few cases adjacent to the spectral lines due to imperfections present in the level 1 data (i.e. this is not introduced by the RVM). We purport that the reconstruction regime effectively removes instrumental defects while retaining the solar information embedded in the Stokes profiles. 

\begin{figure}
\centering
\includegraphics[width=\columnwidth]{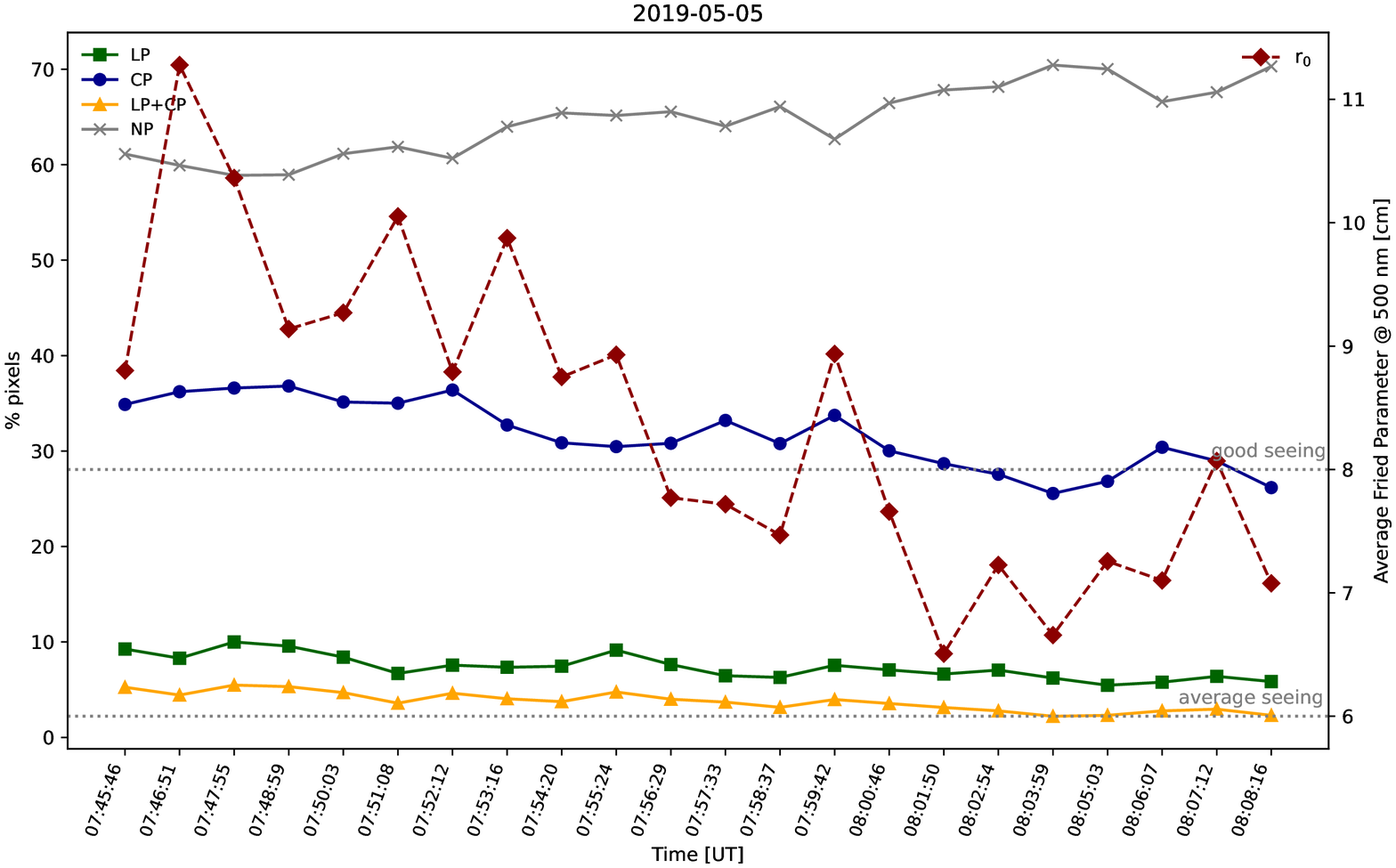}\vspace{0.5cm}
\includegraphics[width=\columnwidth]{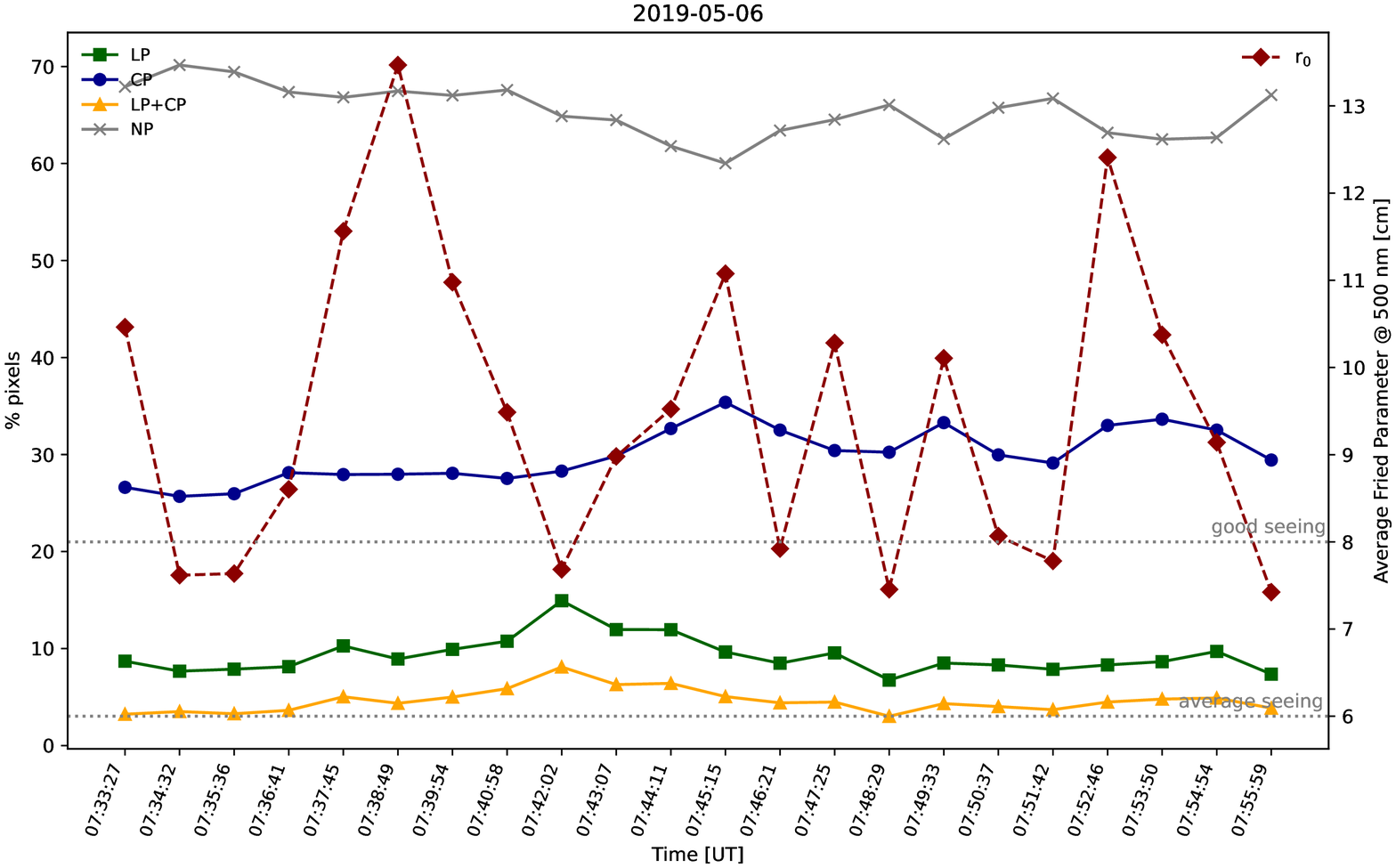}\vspace{0.5cm}
\includegraphics[width=\columnwidth]{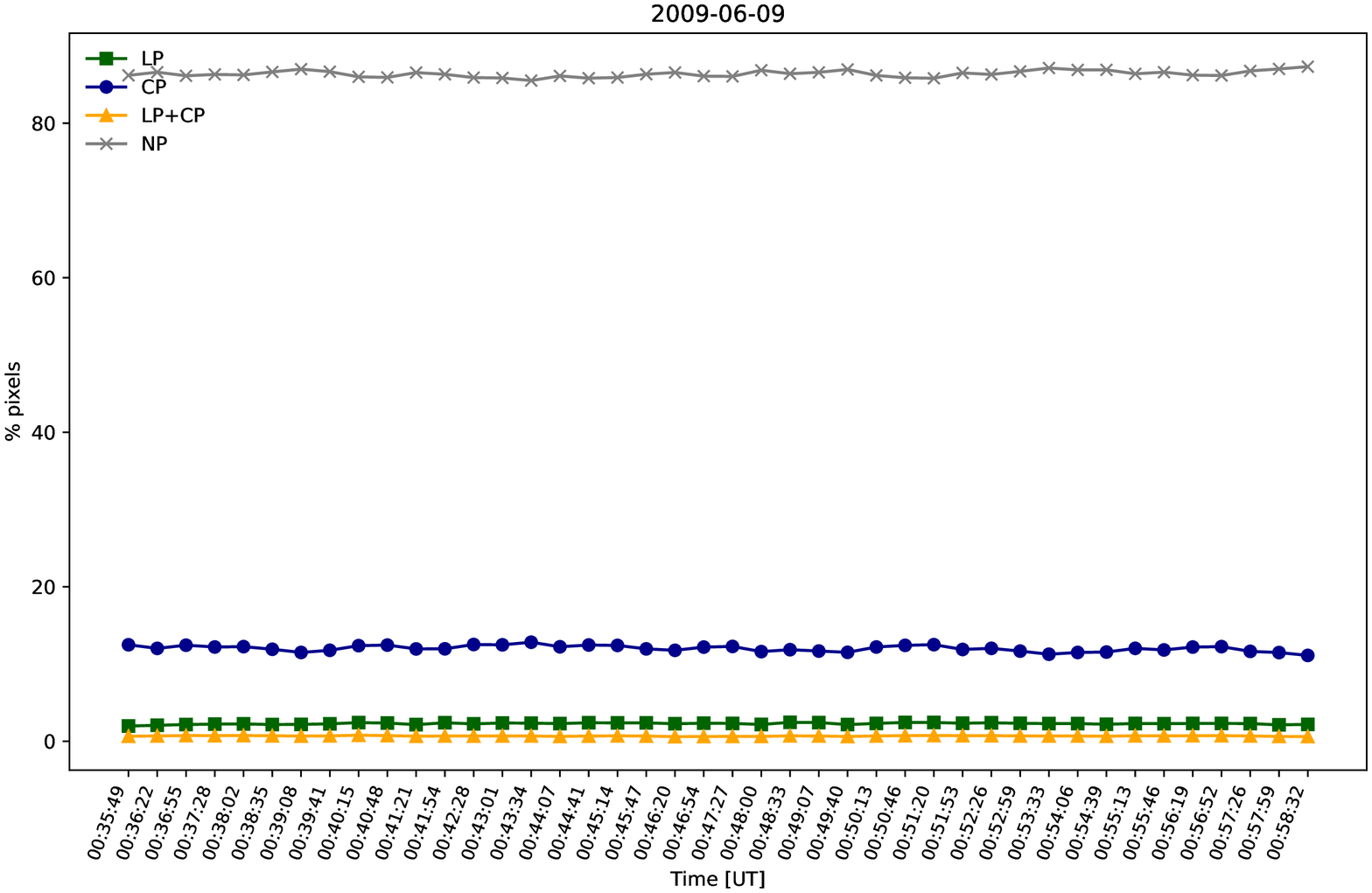}
\caption{Temporal variation in the percentage of pixels with Stokes $Q$ or $U$ (LP), Stokes $V$ (CP), both LP and CP and no polarization (NP) signals thresholded by a 3$\sigma$ value as determined from the standard deviation in the continuum, calculated and plotted for each frame, for the reconstructed GRIS-IFU data taken on the 5 (\textit{upper}) and 6 (\textit{middle}) May. The fried parameter, averaged across all tiles for each frame, is also plotted in red. The maximum r$_0$ value in a single tile was as high as 20 cm, which is considered exceptional. The \textit{lower} plot shows the same, but for the first IMaX scan in Table \ref{table:1}.}
          \label{fig:temporal}%
\end{figure}
\begin{table}
    \caption{As for Table \ref{table:1}, but for the reconstructed GRIS-IFU data thresholded at $\sigma_t$ as measured in the continuum of the original level 1 data.}       
    \label{table:reconstructedstats}      
    \centering                          
    \begin{tabular}{|c|cccc|}        
    \hline
      Date (yyyy-mm-dd) & $\%$LP & $\%$CP & $\%$LP+$\%$CP & $\%$NP  \\   
    \hline 
      2019-05-05 & 7.4 & 31.7 & 3.8 & 64.7 \\
    \hline 
      2019-05-06 & 9.3 & 29.9 & 4.6 & 67.6 \\
      
    \hline
    \end{tabular}
\end{table}

Table \ref{table:reconstructedstats} shows the time-averaged results of the quantification analysis for the GRIS-IFU scans after reconstruction when thresholded at $3\sigma$ as measured in the continuum before reconstruction. The average values for IMaX were $2.3-2.5\%$ LP and $11.2-12.0\%$ CP, while after reconstruction the GRIS-IFU observed $7.4-9.3\%$ LP and $29.9-31.7\%$ CP, on average. The temporal evolution of these quantification statistics for both GRIS-IFU scans, alongside one IMaX scan, on a per-frame basis, is illustrated by Figure \ref{fig:temporal}. The highest values for a single frame throughout both scans was recorded as $14.9\%$ LP at 07:42:02 UT on the $6$ May and $36.8\%$ CP at 07:48:59 UT on the $5$ May. We note that the percentages after reconstruction when thresholded at $3\sigma$ (see Table \ref{table:reconstructedstats}) are close to those before reconstruction when thresholded at between $4\sigma$ and $4.5\sigma$ (see Table \ref{table:1}), which is consistent with what one would expect upon removal of noise. It is also clear from binary maps that the pixels selected by a $4-4.5\sigma$ threshold in the data before reconstruction are indeed the same pixels selected after reconstruction by a $3\sigma$ threshold. Hereafter we refer to the noise threshold as $\sigma_t$ in an effort to avoid confusion, and emphasize that $\sigma_t$ is essentially equivalent to a stringent $4-4.5\sigma$ threshold in terms of selected pixels.

As a final step in the reconstruction process we subtracted the effect of the spectral veil from Stokes $I$. We followed the method of \cite{borrero2016} and iteratively convolved a Gaussian of varying FWHM with the average Stokes $I$ continuum, excluding those pixels with polarization, until a minimum $\chi^2$ was achieved between the convolved profile and the FTS atlas, to measure the spectral point spread function (PSF) of the GREGOR/GRIS-IFU. We found a minimum $\chi^2$ was achieved through convolution with a $\sigma = 73$ m$\AA$ Gaussian and a spectral stray light correction of $11\%$ and $12\%$ for the $5$ and $6$ May scans, respectively. The corrected Stokes $I$ vector is given as
\begin{equation}
    I^{\textrm{cor}} (x,y,\lambda,t) = {\frac{I^{\textrm{obs}}(x,y,\lambda,t)-\nu I_c^{\textrm{obs}}(t)}{(1-\nu)}},
\end{equation}
where $I^{\textrm{obs}}$ is the original observed vector, $\nu$ is the spectral stray light fraction and $I_c^{\textrm{obs}}$ is the averaged continuum level in each frame, t.

\subsection{Inversions}\label{section:inversions}

\begin{table}[b]
\caption{Number of free parameters (nodes) used in given atmospheric variables in each inversion scheme; one where a magnetic atmosphere is embedded with a field-free atmosphere (S1), and another with two magnetic models (S2). The asterisks (*) signify that a parameter is forced to be the same in both models. }       
\label{table:nodes1}      
\centering                          
\begin{tabular}{|c | c | c | c | c |}    
\cline{2-5}
   \multicolumn{1}{c|}{}& \multicolumn{2}{c|}{S1} & \multicolumn{2}{c|}{S2}  \\
\hline                 
  Parameter & Model 1 & Model 2 & Model 1 & Model 2  \\   
\hline 
  T & 5 & 5* & 5 & 5*\\
 
  $v_{mic}$& 1 & 1& 1 & 1 \\

  $v_{LOS}$& 1 & 1& 1 & 1 \\
  
  $v_{mac}$& 1 & 1*& 1 & 1* \\
  
  $B$ & 0 & 1 & 1 & 1\\

  $\gamma$& 0 & 1 & 1 & 1\\

  $\phi$& 0 & 1& 1 & 1\\
  
  $\alpha$& 1 & 1& 1 & 1\\
  
\hline
\end{tabular}
\end{table}

\begin{figure*}
   \centering
    \includegraphics[width=\textwidth]{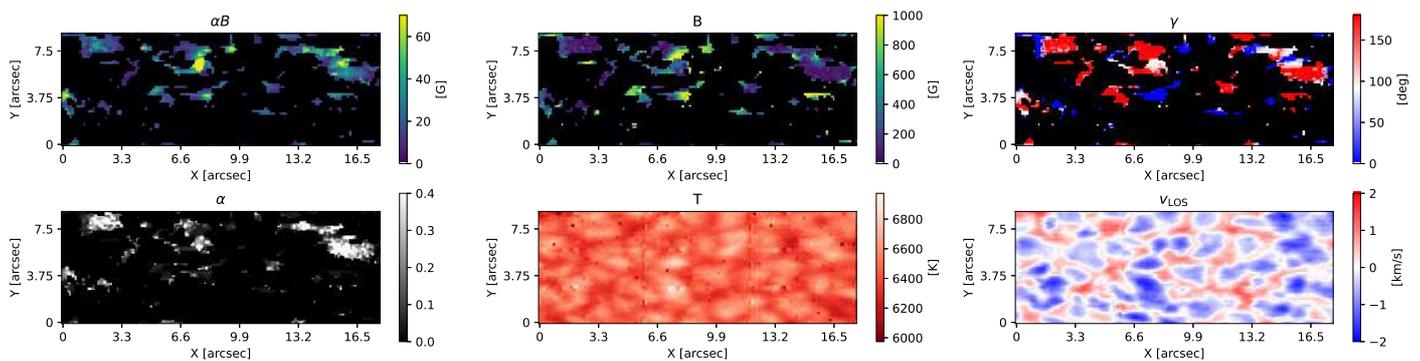}
   \caption{Parameters inferred from S1 SIR inversions of the reconstructed GRIS-IFU data for the frame taken at 07:55:24 UT on the $5$ May, namely, from left-right and top-bottom, $\alpha B$, $B$, $\gamma$, $\alpha$, $T$ at log$\tau_{5000\mathrm{\AA}} = 0$, and $v_{LOS}$. Positive $v_{LOS}$ values represent down-flows, and negative values represent up-flows. A $\gamma$ value of $0^\circ$ represents a vector pointing towards the observer, while $180^\circ$ indicates the opposite. Observables for this frame are shown in Figure \ref{fig:map}.}
              \label{fig:map2}%
    \end{figure*}

\begin{figure}
\centering
\includegraphics[width=.8\columnwidth]{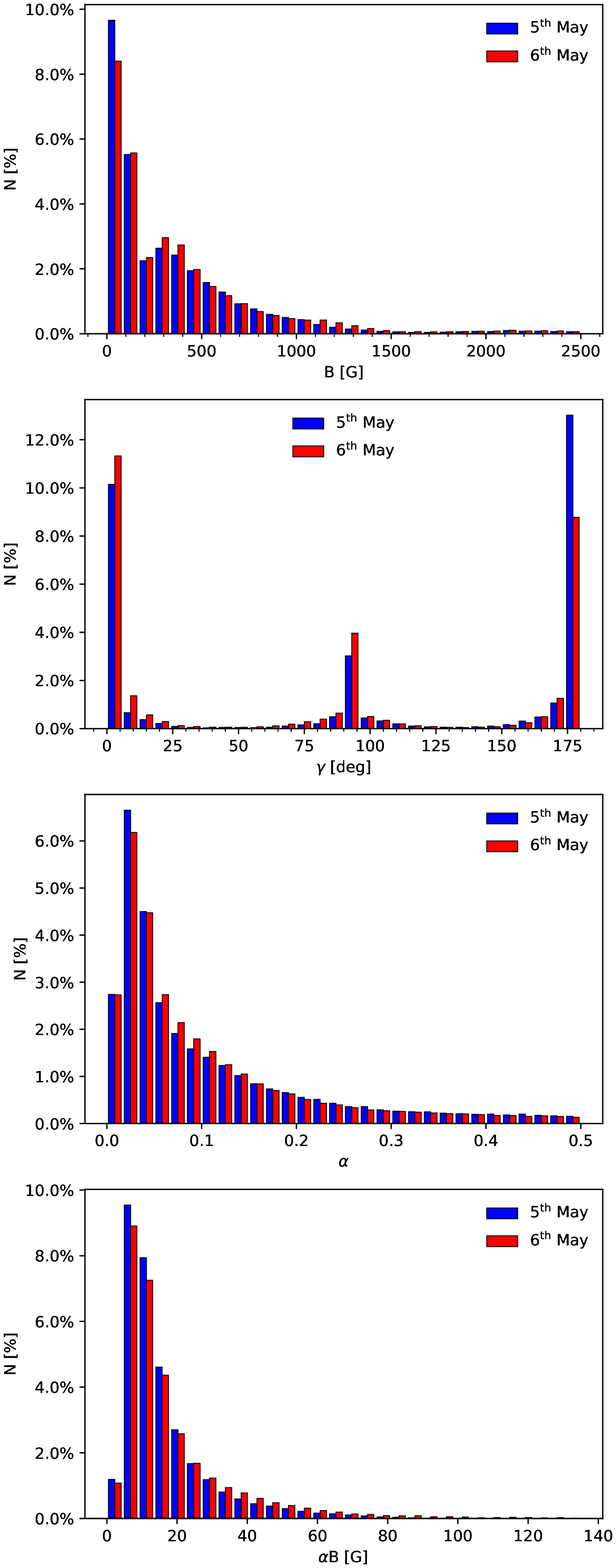}
\caption{Histograms of $B$ (\textit{top row}), $\gamma$ (\textit{second row}), $\alpha$ (\textit{third row}) and $\alpha B$, (\textit{bottom row}) returned from the inversions under S1. The histograms are shown for pixels with maximum amplitude in at least one Stokes vector $\sigma_t$. The \textit{blue} and \textit{red} columns show the distributions for the scans on the $5$ and $6$ May, respectively. The percentages in all histograms are weighted with respect to the total number of pixels, i.e. including those with no polarization.}
          \label{fig:inv_parameters}%
\end{figure}

We analyzed the observations with a least-squares inversion using the Stokes inversion based on response functions (SIR; \citep{SIR}) inversion code to infer the thermodynamic, kinematic and magnetic properties of the atmosphere from the observed Stokes spectra. The inversion code uses an initial guess model, which describes each of the parameters characterizing the stratification of the atmosphere in optical depth - namely temperature, $T$, electron pressure, gas density, gas pressure, microturbulent velocity, $v_{\mathrm{mic}}$, unsigned magnetic field strength, $B$, LOS velocity, $v_{\mathrm{LOS}}$, inclination of the magnetic field with respect to the observer's LOS, $\gamma$, and azimuth of the magnetic field in the plane perpendicular to the observer's LOS, $\phi$ - to compute a synthetic Stokes vector by solving the radiative transfer equation under the assumption of local thermodynamic equilibrium (LTE). This synthetic vector is then compared to the observed spectra through a $\chi^2$ merit function. A Levenberg-Marquardt algorithm and singular value decomposition method is then employed to perturb the guess model by an amount at a variable number of `nodes' placed at given optical depths in a number of the atmospheric parameters, with the final depth stratification obtained by cubic-splines or linear interpolation between the nodes. The process of solving the radiative transfer equation and perturbing the model is then iterated until a minimum $\chi^2$ is achieved. 

We developed a program, based on a code used and developed by \cite{borrero2016} to run SIR inversions, allowing us to repeat the inversion procedure $15$ times per pixel with randomized model parameters for each Stokes vector (in $\gamma, \phi, B, v_{\mathrm{LOS}}$) in order to reduce the probability of converging at a solution that is located in a local, as opposed to global, minimum in the $\chi^2$ hyper-surface. We used a FALC model \citep{font2006} to provide the initial atmospheric stratifications for the remaining model parameters. The model that produces the minimum $\chi^2$ value is taken as the final solution. Table \ref{table:atomicdata} shows the atomic data used in the inversion for the five Fe $\textsc{I}$ lines in the observed spectral window. We invert the profiles as reconstructed by the full process described in section \ref{section:reconstruction}. In order to prevent the over-interpretation of noisy Stokes profiles, we measure the average $1\sigma$ noise level in the continuum of the original level 1 data and any Stokes $Q$, $U$, or $V$ profile whose maximum amplitude does not exceed $\sigma_t$ is set to zero. This is an attempt at preventing the SIR code from inserting, in particular, linear polarization signals where we cannot say with confidence that we have a real signal.

In our first inversion scheme, scheme 1 (S1), we inverted the data by modelling each resolution element as comprising a magnetic and non-magnetic plasma. Table \ref{table:nodes1} describes the nodes used in this scheme. The magnetic parameters were assumed to be constant with height in the atmosphere, and $\alpha$ was also inverted. Temperature was allowed to vary at 5 nodes but forced to be the same for both models. Macroturbulent velocity, $v_{\mathrm{mac}}$, was also forced to be the same in both models. We do not include a spectral PSF as an input, as we do not empirically know GREGOR's spectral PSF and in any case in SIR a Gaussian is convolved through $v_{\mathrm{mac}}$ in Fourier space in the same way as an estimated spectral PSF would be. Therefore, we instead allowed $v_{\mathrm{mac}}$ to encapsulate the spectral PSF as a free parameter. Initializations in $\gamma, \phi, B, v_{\mathrm{mic}}$ and $v_{\mathrm{LOS}}$ were randomized, but the filling factor of both models was always initialized as $0.5$.  This model is expected to be able to produce synthetic profiles with good fits to the data for those with low area and amplitude asymmetries in Stokes $V$ profiles with two lobes, and by extension low velocity gradients, while constraining $\alpha$. The model is also only expected to fit Stokes $Q$ and $U$ profiles that are compatible with the Stokes $V$ profiles and, indeed, polarized vectors that are compatible with Stokes $I$. However, as the velocity is not forced to be the same in both models, SIR should be capable of producing synthetic polarized vectors that are Doppler shifted to a different extent than the corresponding Stokes $I$ profile. In cases where the model is a good approximation, $B$ and $\alpha$ should be constrained. For the weakest fields, we do not expect to be able to precisely determine $\gamma$ but only whether the vector may be considered to be predominantly longitudinal or transverse in nature. We did not expect to be able to constrain $\phi$ for the vast majority of pixels; although it may be constrained for those with strong Stokes $Q$ and $U$ profiles, but even in these cases we did not make an attempt at disambiguation. We present statistical results determined from S1 in section \ref{section:statisticalanalysis}. In scheme 2 (S2), we considered inversions where both models are magnetic. Additionally, in this scheme, we introduced a quiet Sun averaged unpolarized stray light profile, $\mathrm{I}_\mathrm{stray}$, that is added to the synthetic profiles. We assume a stray light fraction of $30\%$, following the estimation by \cite{borrero2016} for GRIS/GREGOR. We applied S2 only to select pixels of interest in section \ref{section:regionsofinterest}, in an attempt to model multi-lobed polarization profiles. As S2 has a larger number of free parameters than S1, we repeated each inversion $150$ times. Appendix \ref{section:RFs} provides further justification for our choice of inversion schemes in the context of response functions.

\subsubsection{Statistical analysis}\label{section:statisticalanalysis}
Figure \ref{fig:map2} shows a map of the parameters retrieved from the inversion under S1 for one frame of the dataset taken on $5$ May at 07:55:24 UT. The temperature map strongly resembles the Stokes $I$ map, with strong positive correlation between these two parameters as expected. Further, there is a strong negative correlation between Stokes $I$ and $v_{\mathrm{LOS}}$.  From the $\gamma$ maps, it is clear that the polarity of the magnetic field is well retrieved from the circular polarization. Naturally, inclined fields are found in regions with strong linear polarization. Histograms of the magnetic parameters are shown in Figure \ref{fig:inv_parameters} for all inverted frames. There are two distinct populations; a weaker but highly populated group of weak fields, and a stronger but lower populated group of kilo-Gauss fields. There is a small peak in the $B$ distributions of both scans at {\raise.17ex\hbox{$\scriptstyle\mathtt{\sim}$}} $300$ G. The vast majority of the $\alpha$ values are small.

We define the unsigned average magnetic field strength,
\begin{equation}
    \mu_B = \frac{\sum\limits_{i=1}^{N} |B_i| }{N},
\end{equation}
where $N$ is the total number of pixels with maximum amplitude greater than $\sigma_t$ in at least one polarized Stokes parameter across the $15648.52$ $\AA$ line. The mean transverse and longitudinal (unsigned) components, $\mu_\perp$ and $\mu_\parallel$ respectively, are similarly defined by replacing $|B_i|$ with $B_\perp = |B_i \sin{\gamma_i}|$ and $B_\parallel = |B_i \cos{\gamma_i}|$, respectively. However, when calculating these two parameters, we consider only profiles that had Stokes $Q$ or $U$ $> \sigma_t$ in the former case, and Stokes $V$ $> \sigma_t$ in the latter case. Here we can explicitly make the assumption that, as the observations were recorded at disk-centre, the longitudinal component of the magnetic field with respect to the observer is equal to the vertical component of the field with respect to the solar surface (i.e. parallel to the solar normal), $B_z = B_\parallel$, and similarly for the transverse component of the field and the horizontal component field (i.e. perpendicular to the solar normal), $B_h = B_\perp$. To be clear, we define the unsigned magnetic field strength, $B$, and its unsigned longitudinal and transverse components, $B_\parallel$ and $B_\perp$, respectively. The quantity $\alpha B$ is thus the unsigned magnetic flux density, while $\alpha B_\perp$ and $\alpha B_\parallel$ are the unsigned longitudinal and transverse magnetic flux densities, respectively. The results for each scan is shown in Table \ref{table:means_05}, along with corresponding standard deviations in each mean. Clearly, deriving meaning from the means of such distributions is problematic as the standard deviations are so high relative to the values. Further, the mean is sensitive to anomalous high or low values of $B$ in pixels that have poorly converged. A better parameter to consider, therefore, is the median, which is also shown in Table \ref{table:means_05}, as the median is less sensitive to anomalous values. We may also consider the ratio of the medians of $B_\perp$ to $B_\parallel$, which is returned as $1.8-2.1$. The mean and median of $\alpha B$ is $16$ G and $12$ G, respectively, for the $5$ May scan, and $19$ G and $13$ G, respectively, for the $6$ May scan.

\begin{table}
\caption{Unsigned average and median magnetic field strength, and its horizontal and vertical components. The mean and median $B$ is measured across the entire population of values for the datasets taken on the $5$ and $6$ May 2019, but when computing the corresponding $B_\perp$ and $B_\parallel$ component we consider only profiles that had Stokes $Q$ or $U$ $> \sigma_t$ in the former case, and Stokes $V$ $> \sigma_t$ in the latter case. The mean values are shown accompanied by the standard deviations in brackets.}       
\label{table:means_05}      
\centering                          
\begin{tabular}{|c | c | c | c| c|}

\cline{2-5}
   \multicolumn{1}{c|}{}& \multicolumn{2}{c|}{$5$ May 2019} & \multicolumn{2}{c|}{$6$ May 2019}  \\
\cline{2-5}
   \multicolumn{1}{c|}{}& \multicolumn{1}{c|}{mean} & \multicolumn{1}{c|}{median} &
    \multicolumn{1}{c|}{mean} & \multicolumn{1}{c|}{median}\\

\hline
  $B$ & 356 ($\pm$437) G & 200 G & 385 ($\pm$467) G & 240 G \\
 
  $B_\perp$ & 563($\pm$730) G & 263 G & 544($\pm$713) G & 267 G \\

  $B_\parallel$ & 269 ($\pm$286) G & 131 G & 280 ($\pm$301) G & 145 G\\
  
\hline
\end{tabular}
\end{table}
\begin{table}
    \caption{Percentage of profiles with inclinations in given ranges as determined by inversions under S1 for the scan taken on $5$ (third column) and $6$ May 2019 (fourth column). The percentage is shown for those pixels with a maximum amplitude above $\sigma_t$ in at least one Stokes vector, and are computed relative to the total number of profiles in each population.}       
    \label{table:inclin05}      
    \centering                          
    \begin{tabular}{|c |c | c | c|}        
    \hline
      classification & range [$^\circ$] & $5$ $[\%]$ & $6$ $[\%]$\\   
    \hline 
      highly vertical&$\gamma < 16$ & 34.2 & 40.7\\
     
      highly vertical&$\gamma > 164$& 44.8 & 32.5 \\
    
      intermediate&$15 < \gamma < 75$& 2.8 & 4.3 \\
      
      intermediate&$105 < \gamma < 165$& 4.4 & 4.1 \\
      
      highly inclined&$74 < \gamma < 106$ & 13.8 & 18.4\\
      
    \hline
    \end{tabular}
\end{table}

\begin{figure*}
    \centering
\includegraphics[width=\textwidth]{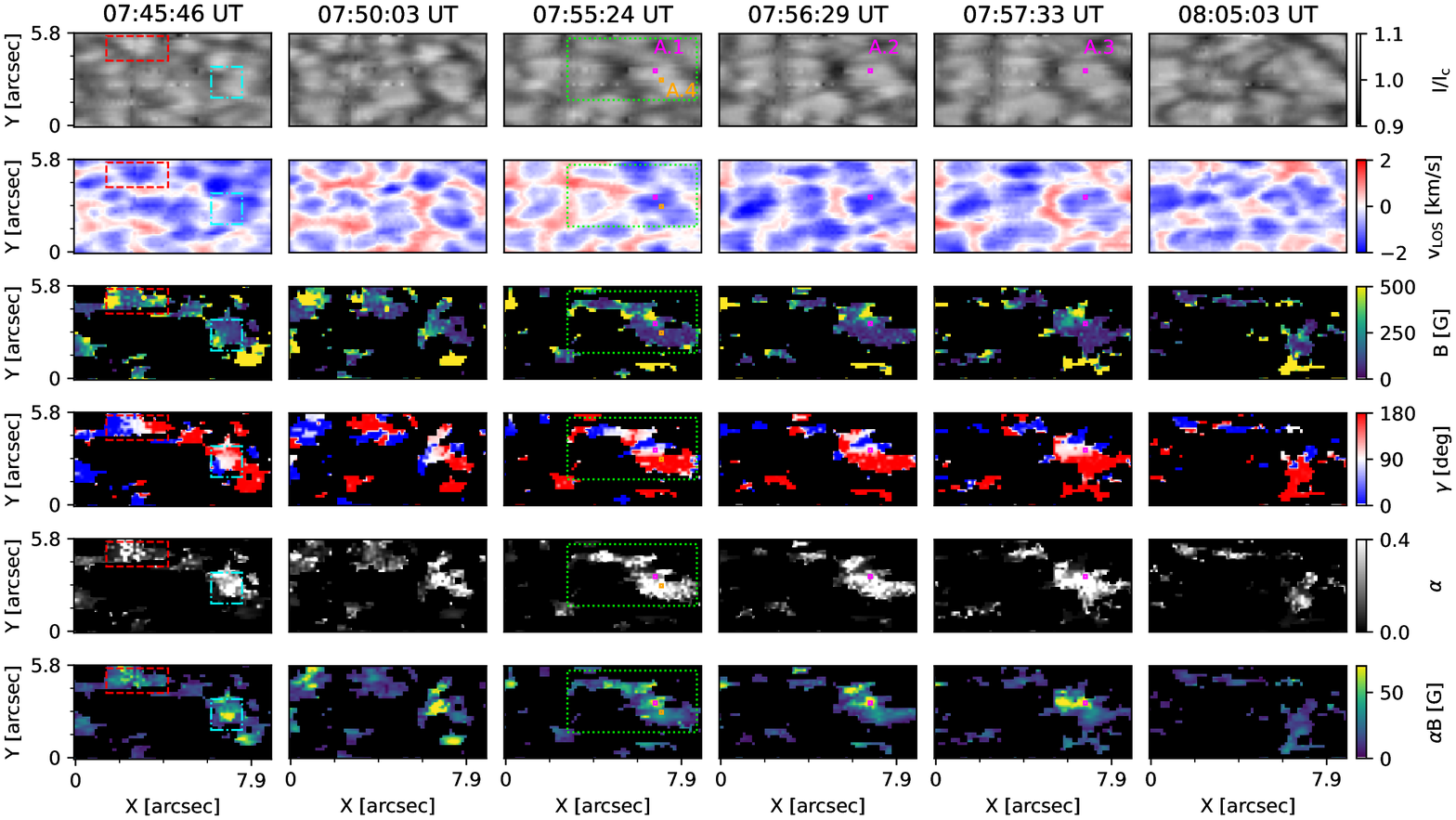}
\includegraphics[width=\textwidth]{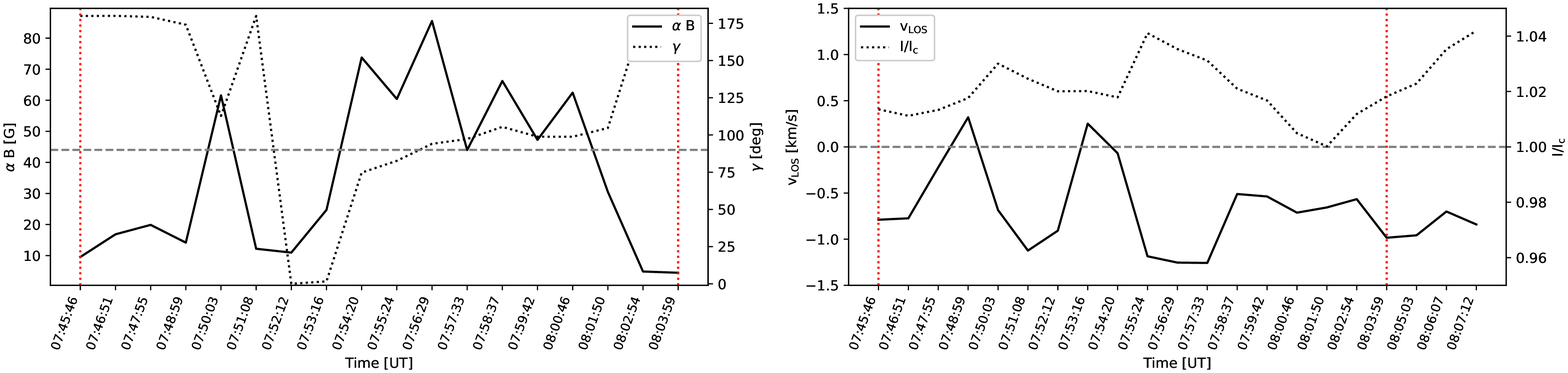}
\caption{ROI A: A region with complex loop-like structures, observed in the $5$ May scan, with clear opposite polarity longitudinal and transverse components present in every frame. \textit{Upper panels:} First row showing the Stokes $I$ normalized continuum, with subsequent rows showing, in descending order, as retrieved from S1 inversions: $v_{\mathrm{LOS}}$, $B$, $\gamma$, $\alpha$, and $\alpha B$. Maps of $B$ saturate at $500$ G so weaker fields are more easily visible. The columns show a selection of subsequent frames. \textit{Bottom panels}: Temporal evolution of the following quantities are shown for the pixel whose spatial and temporal location in ROI A is highlighted with a solid magenta square outline above (i.e. the pixels A.1-3): $\alpha B$, $\gamma$ (\textit{left}) and $v_{\mathrm{LOS}}$, $I/I_{\mathrm{c}}$ (\textit{right}) as derived from S1 inversions. The \textit{horizontal dashed, grey} lines indicate the position of $90^\circ$ on the left panel and $0$ km/s on the right panel. The \textit{vertical dotted, red} lines are a visual aid for the time-stamps of the start and end-point of the left panel.}
    \label{fig:ROIA}
\end{figure*}

The distributions of inclinations shown in Figure \ref{fig:inv_parameters} has five clear populations; Table \ref{table:inclin05} illustrates how the field can be classified in terms of highly vertical fields, highly inclined fields and those with no clear preference in direction. The large peaks in the distributions at $0,90$ and $180^\circ$ are a consequence of having set any Stokes parameter with signal $< \sigma_t$ to zero. We tested the inversions without setting these noisy Stokes vectors to zero, and the result was that the inversion scheme inserted polarization signals into the noise and therefore a much higher population of inclinations with no clear preference in orientation was returned. For the $5$ May scan, the highly vertical populations dominate, together accounting for $78.9\%$ of profiles with signal greater than $\sigma_t$. While the proportion that has highly inclined fields is relatively low, at $13.9\%$, the total proportion of fields with a clear transverse component is $21.1\%$. For the $6$ May scan, the highly vertical populations are marginally lower but nevertheless dominant, together accounting for $73.2\%$ of profiles with signal greater than $\sigma_t$.  Indeed, the proportion that has highly inclined fields is relatively high in this scan, at $18.4\%$, while the total proportion of fields with a clear transverse component is $26.8\%$. We additionally observe a clear polarity imbalance in the inclinations, as was also reported by \cite{marian} with GRIS, however the polarity is imbalanced in opposite directions in each of the scans. Given the small FOV, it is impossible to infer whether this has any significance due to these instrumental restrictions.

\subsubsection{Regions of interest}\label{section:regionsofinterest}

We define, curate, and present a number of ROIs, displaying small-scale magnetic features, and aim to track their temporal evolution. We are particularly concerned with ROIs that show linear polarization signals, as in these regions we have the greatest chance of attempting to fully characterize the magnetic vector. We use S1 inversions to constrain the broad atmospheric conditions across the ROI and its immediate surroundings at each time-step. Precisely tracking the magnetic properties of such small-scale internetwork features is a very difficult task; the polarization amplitudes are often so close to the $\sigma_t$ noise level that Stokes vectors often fall below the threshold between frames. However, we are able to track the thermodynamic and kinematic properties for every frame.  We attempt to define, within the constraints of the cadence, spatial resolution and FOV, each feature's life-time, location with respect to the granules and IGLs and its fate in terms of emergence and submergence during appearance and disappearance. We define a `loop-like' structure as one in which circular polarization, of opposite polarities, flanks linear polarization, such that the linear polarization is located in a PIL. We also define an LPF as a small element of inclined fields, appearing in isolation or close to largely mono-polarity circular polarization. One should examine maps of $\gamma$ to locate these structures when referred to hereafter. One should also examine the $v_{\mathrm{LOS}}$ maps to discern the location of a given structure in the granulation.

\begin{figure*}
    \centering
    \includegraphics[width=\textwidth]{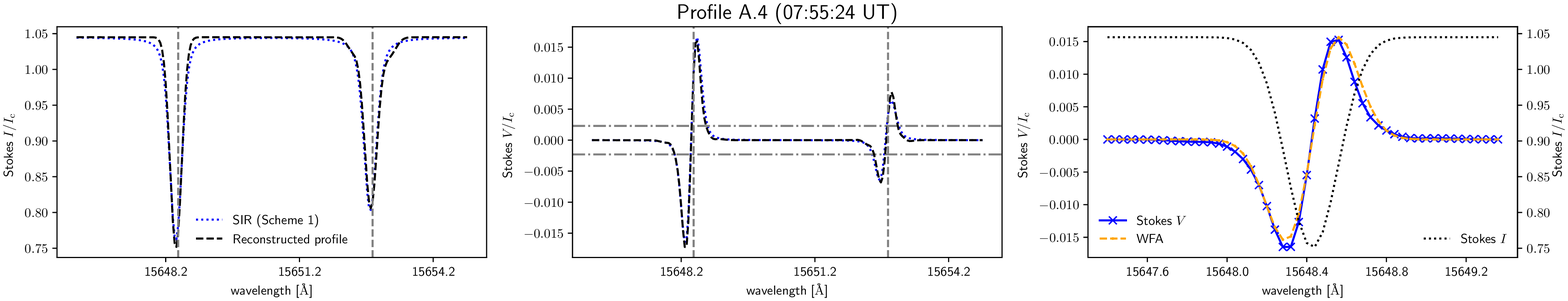}
    \caption{Example profile with its WFA estimate shown next to its the synthetic fit as returned from S1 inversions. \textit{Left and middle panels:} Reconstructed Stokes profiles (\textit{black, dashed line}) from a pixel whose temporal and spatial location is indicated by the orange square outline in Figure \ref{fig:ROIA}. Only Stokes $V$ in this pixel had a maximum amplitude greater than $\sigma_t$. The \textit{blue, dotted line} shows the minimum $\chi^2$ synthetic Stokes vector produced by SIR under S1 inversions. The horizontal (\textit{dot-dashed}) lines show the $\sigma_t$ noise threshold for the Stokes $V$, while the vertical (\textit{dashed}) lines denote the rest wavelengths of each spectral line. The retrieved parameters were: $B = 65$ G, $\gamma = 179^\circ$, $\alpha = 0.61$, $v_{\mathrm{LOS}} = -1.06$ km/s.
    \textit{Right panel:} Stokes $I$ (\textit{black, dotted line}) and Stokes $V$ (\textit{blue line with markers}) profile over-plotted with the WFA fit obtained from the derivative of Stokes $I$ (\textit{orange, dashed line}).}
    \label{fig:WFA}
\end{figure*}

\begin{figure*}
    \centering
    \includegraphics[width=\textwidth]{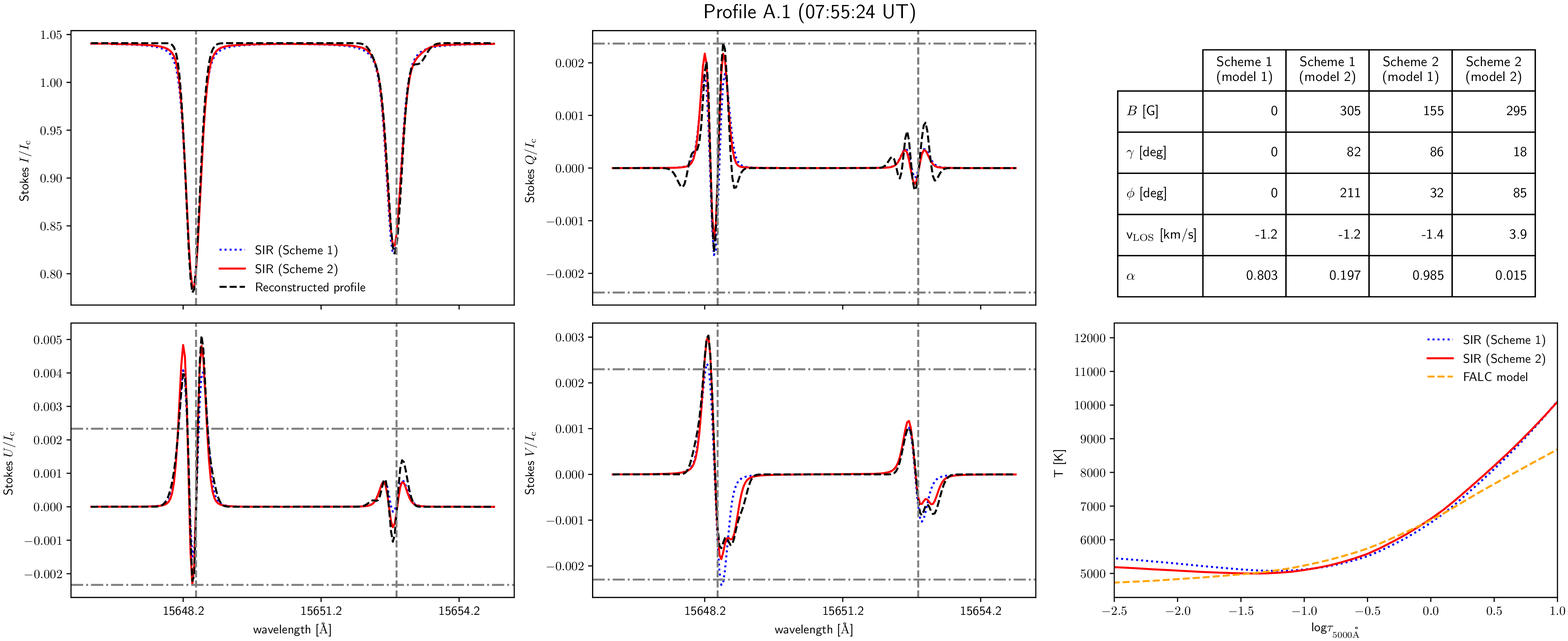}\vspace{0.1cm}
    \includegraphics[width=\textwidth]{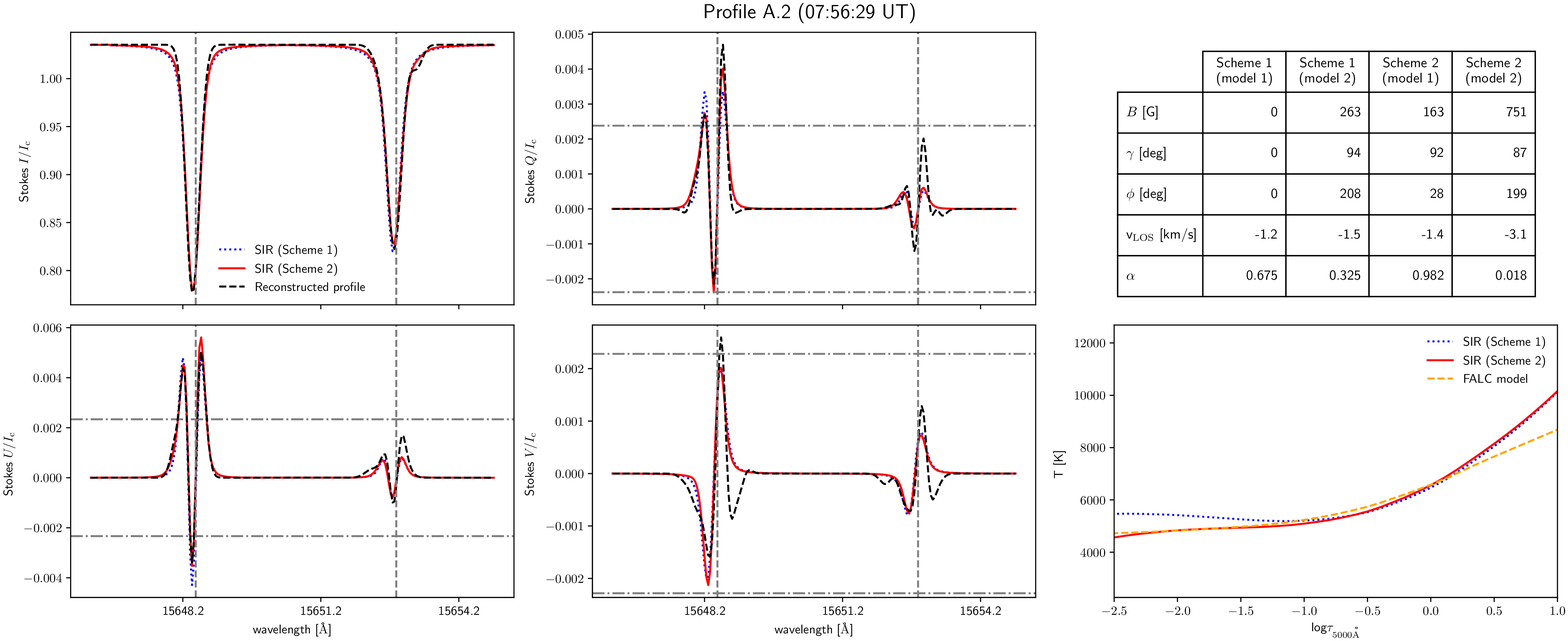}\vspace{0.1cm}
    \includegraphics[width=\textwidth]{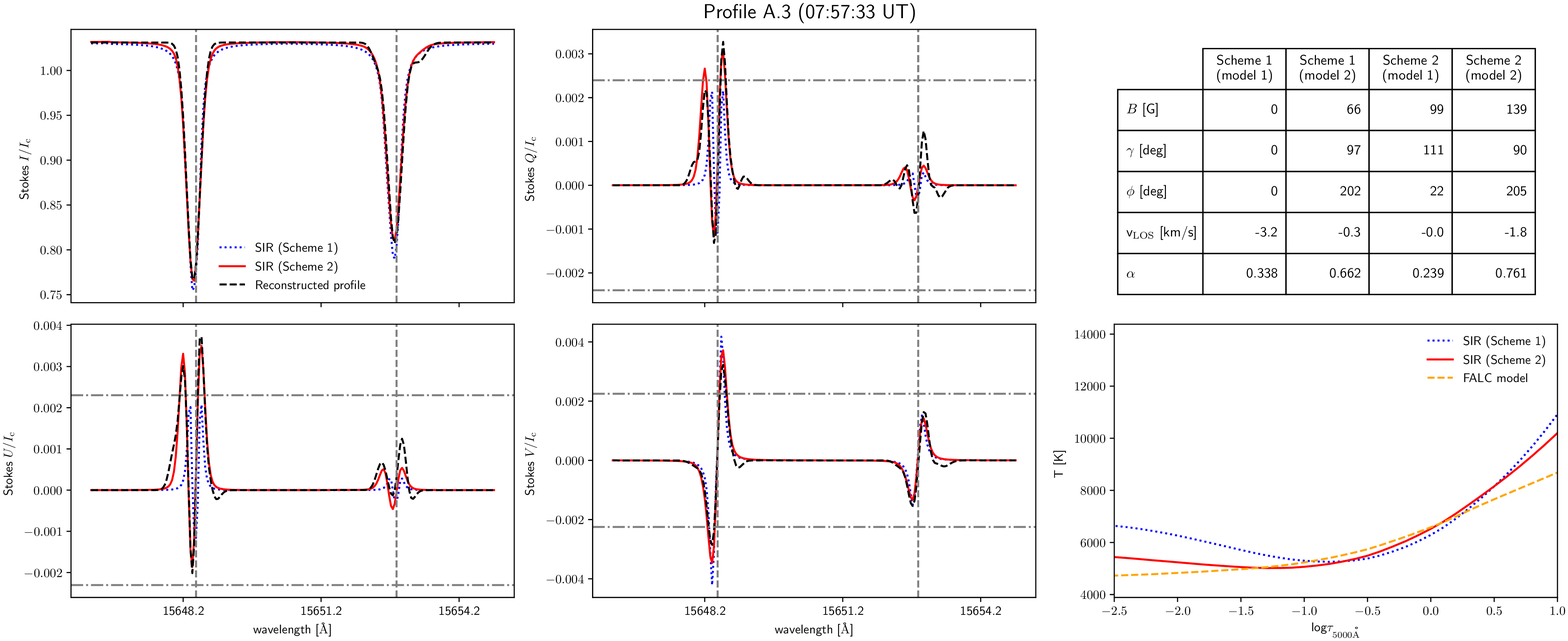}
    \caption{Reconstructed full Stokes vector is shown for the $15648.52/15652.87 \AA$ line pair in the left four panels, along with the S1 and S2 synthetic profiles derived from the SIR inversions, for the pixel whose spatial location in ROI A is highlighted with magenta square outline in Figure \ref{fig:ROIA}, for three frames. The horizontal (\textit{dot-dashed}) lines show the $\sigma_t$ noise thresholds for the polarized Stokes vectors, while the vertical (\textit{dashed}) lines denote the rest wavelengths of each spectral line. On the right, the  retrieved atmospheric parameters for each scheme is shown in the table, while the temperature as a function of optical depth is shown in the lower right plot with the original FALC input model.}
    \label{fig:ROIA_pix}
\end{figure*}

\begin{figure*}
    \centering
\includegraphics[width=\textwidth]{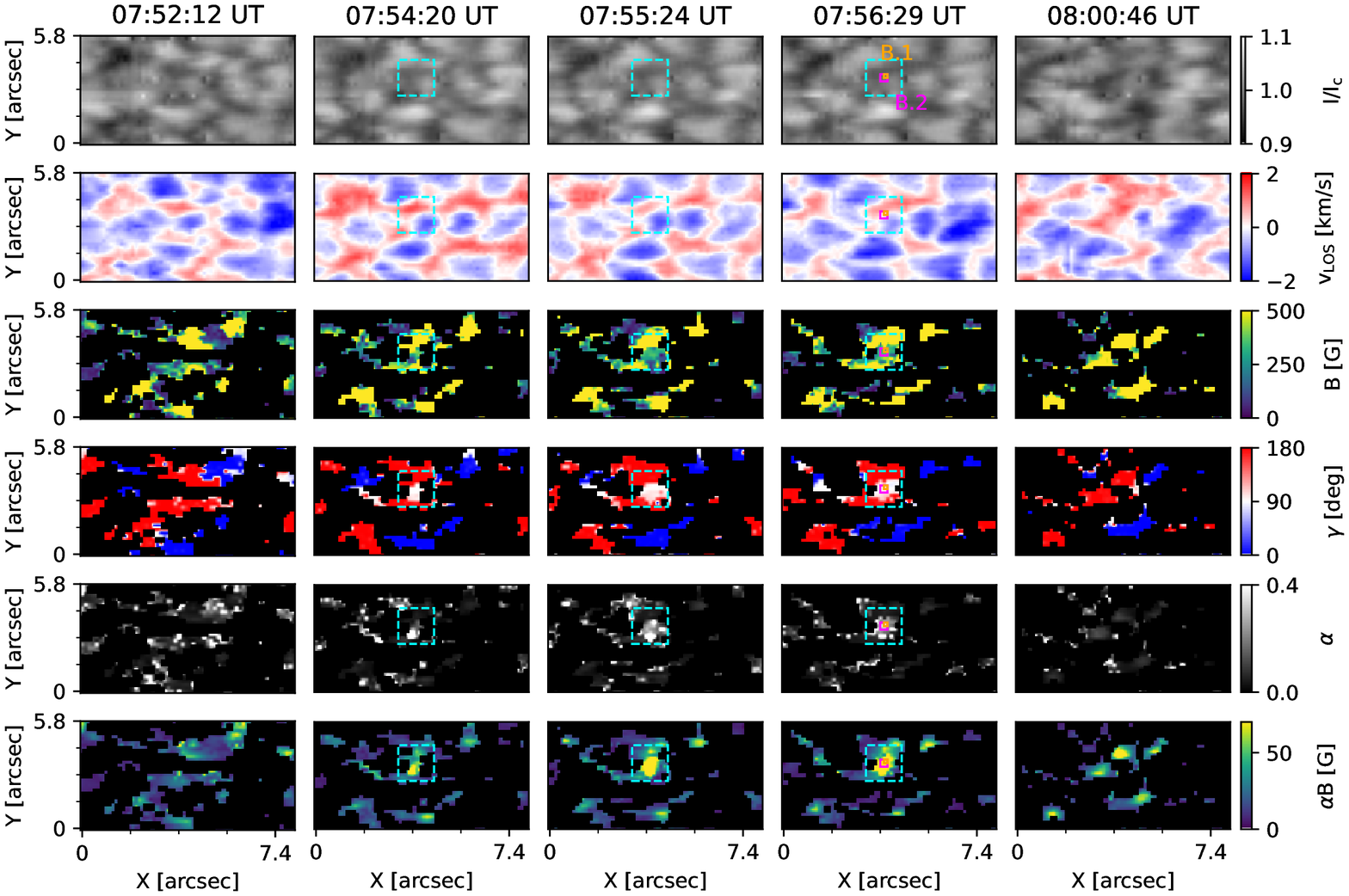}\vspace{.1cm}
\includegraphics[width=\textwidth]{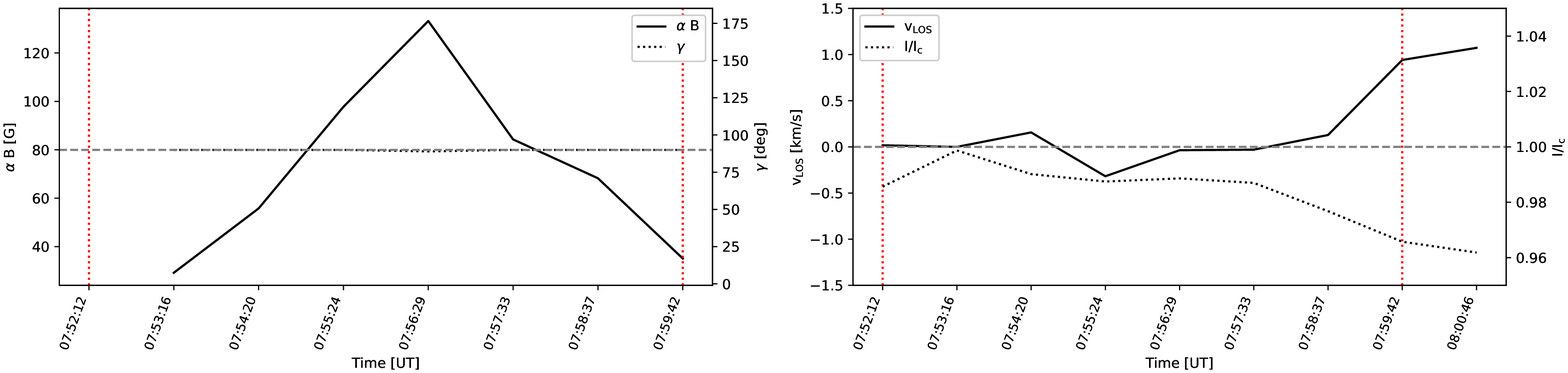}
\caption{ROI B: As in Figure \ref{fig:ROIA}, but for a short-lived linear polarization feature (LPF) observed in the $5$ May scan. The \textit{lower} plots show the temporal evolution of the atmospheric parameters retrieved from S1 inversions for the pixel labelled B.1. B.2 is the resultant profile after binning across the four pixels (including B.1) enclosed by the magneta square.}
    \label{fig:ROIB}
\end{figure*}

\begin{figure*}
    \centering
    \includegraphics[width=\textwidth]{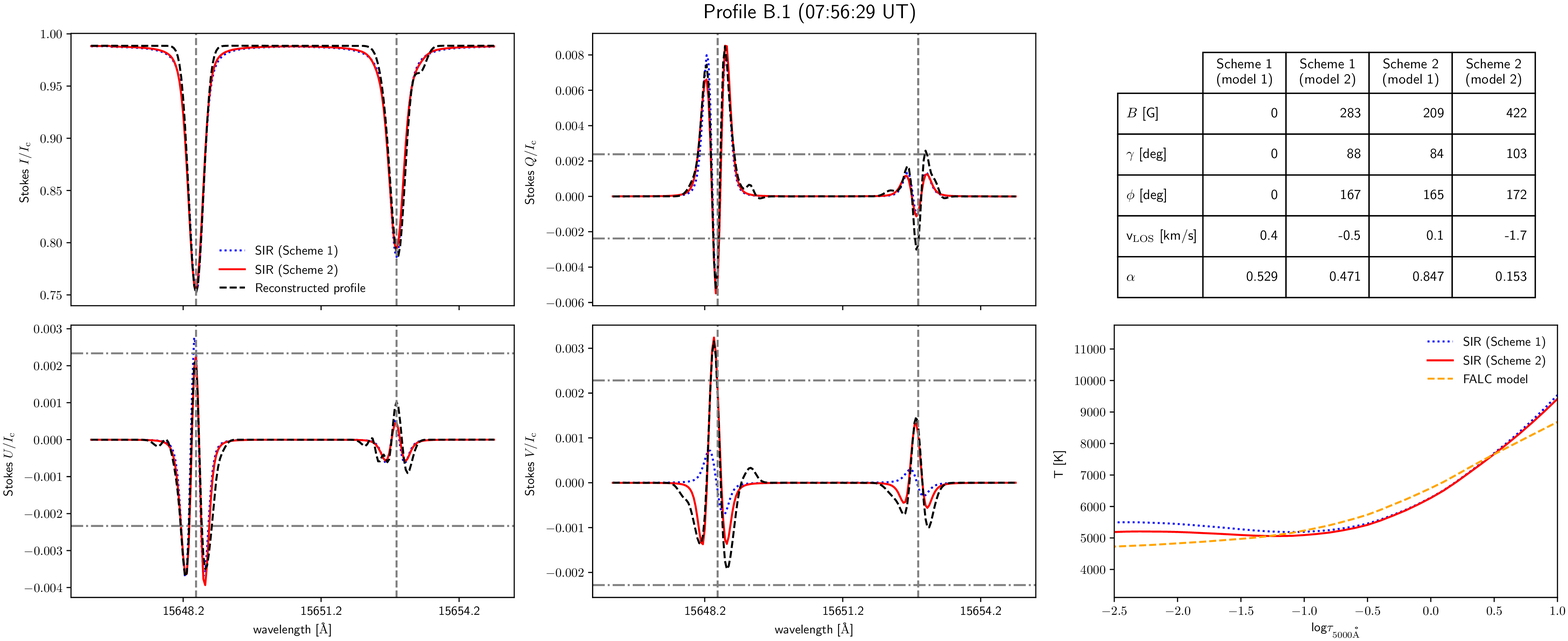}\vspace{0.1cm}
    \includegraphics[width=\textwidth]{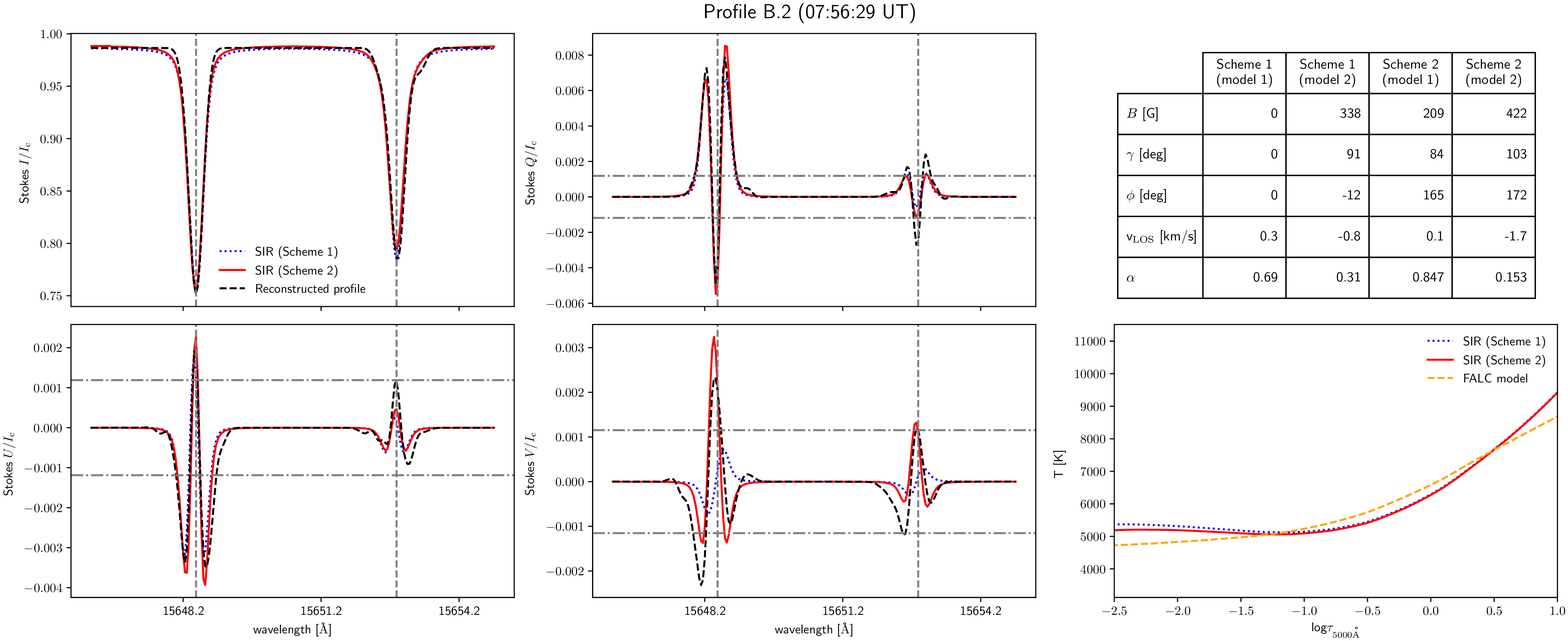}
    \caption{Profiles B.1 and B.2: As in Figure \ref{fig:ROIA_pix}, but for the pixels whose temporal and spatial location in ROI B is highlighted in Figure \ref{fig:ROIB}. B.2 is the resultant profile after binning across the four pixels (including B.1) enclosed by the magneta square in Figure \ref{fig:ROIB}.}
    \label{fig:ROIB_pix}
\end{figure*}

Figure \ref{fig:ROIA} shows a selection of frames from ROI A, observed in the $5$ May scan. This is a very complex feature, with many regions of opposite polarities in close contact. Thus, at any given time, this feature has many sites where cancellation could have potentially taken place. It is also a feature that contains an abundance of inclined fields. We cannot observe the appearance of this feature, due to artefacts visible in Stokes $I$ prior to the frames shown. At 07:45:46 UT, we observe a clear loop-like structure in the top left of the FOV (enclosed by the dashed, red square in Figure \ref{fig:ROIA}) and right of centre we also observe an LPF surrounded by a largely but not exclusively mono-polar field (enclosed by the dot-dashed, cyan square in Figure \ref{fig:ROIA}). In the former case, linear polarization, clearly located in a granule, seems to separate opposite polarity longitudinal flux, located in neighbouring IGLs, and by 07:50:03 UT the linear polarization has disappeared. By 07:55:24 UT, we observe that the top left and right-of-centre structures appear to have formed a continuous magnetic structure across the surface of a granule (enclosed by the dotted, green square in Figure \ref{fig:ROIA}).  By 08:05:03 UT the linear polarization has largely disappeared and the typical $\alpha$ values have greatly diminished compared to other frames, and, thus, so has the $\alpha B$.

Figure \ref{fig:ROIA} also shows the evolution in the magnetic and kinematic properties of the profile, whose location is marked and labelled as A.1-3, between 07:45:46 and 08:03:59 UT. There are two polarity switches; one between 07:51:08 and 07:52:12 UT and another between 07:55:24 and 07:56:29 UT. The $\alpha B$ is greatest when the field is inclined. This pixel is clearly located in a granule, and as such has a brightness typically brighter than the average continuum, but the pixel becomes progressively darker during the presence of significant $\alpha B$ (when the field is inclined) between 07:54:20 and 08:01:50 UT before becoming quickly brighter again once $\alpha B$ disappears (when the field is no longer inclined). The pixel generally experiences an up-flow throughout this time-series, but the pixel experiences changes in its kinematics at key points; the change in $\gamma$ at 07:48:59 UT occurs during a down-flow, and the polarity switch between 07:51:08 and 07:52:12 UT is preceded by an up-flow but followed immediately by a down-flow. There is a clear anti-correlation between $I/I_{\mathrm{c}}$ and $v_{\mathrm{LOS}}$ and these quantities appear to vary periodically on a timescale consistent with $5$ min granular oscillations, as was also reported by \cite{IMaX_marian_2011}.

In appendix \ref{section:WFA} an introduction to the weak field approximation (WFA) is provided in the context of the $15648.52$ $\AA$ line. We can consider the WFA to obtain a lower limit on $\alpha B_\parallel$, and compare this value to that provided by the inversions using the retrieved $\alpha$, $\gamma$ and $B$ values. Figure \ref{fig:WFA} shows an example Stokes vector with the corresponding synthetic fit produced by S1 inversions, taken from the 07:55:24 UT frame in ROI A, and whose spatial and temporal location is marked and labelled as A.4 in Figure \ref{fig:ROIA}. The profile has only circular polarization above the $\sigma_t$ noise threshold, and is very symmetric. The inversion returned $B$ as $65$ G - firmly in the weak field regime - and $\gamma$ as $179^\circ$. The $\alpha B_\parallel$ value of the WFA estimate was returned as $-40.4$ G, which translates to $B_\parallel = 66.0$ G for $\alpha = 0.61$ - within {\raise.17ex\hbox{$\scriptstyle\mathtt{\sim}$}}$1\%$ of the value retrieved by the inversion.

\begin{figure*}
    \centering
\includegraphics[width=\textwidth]{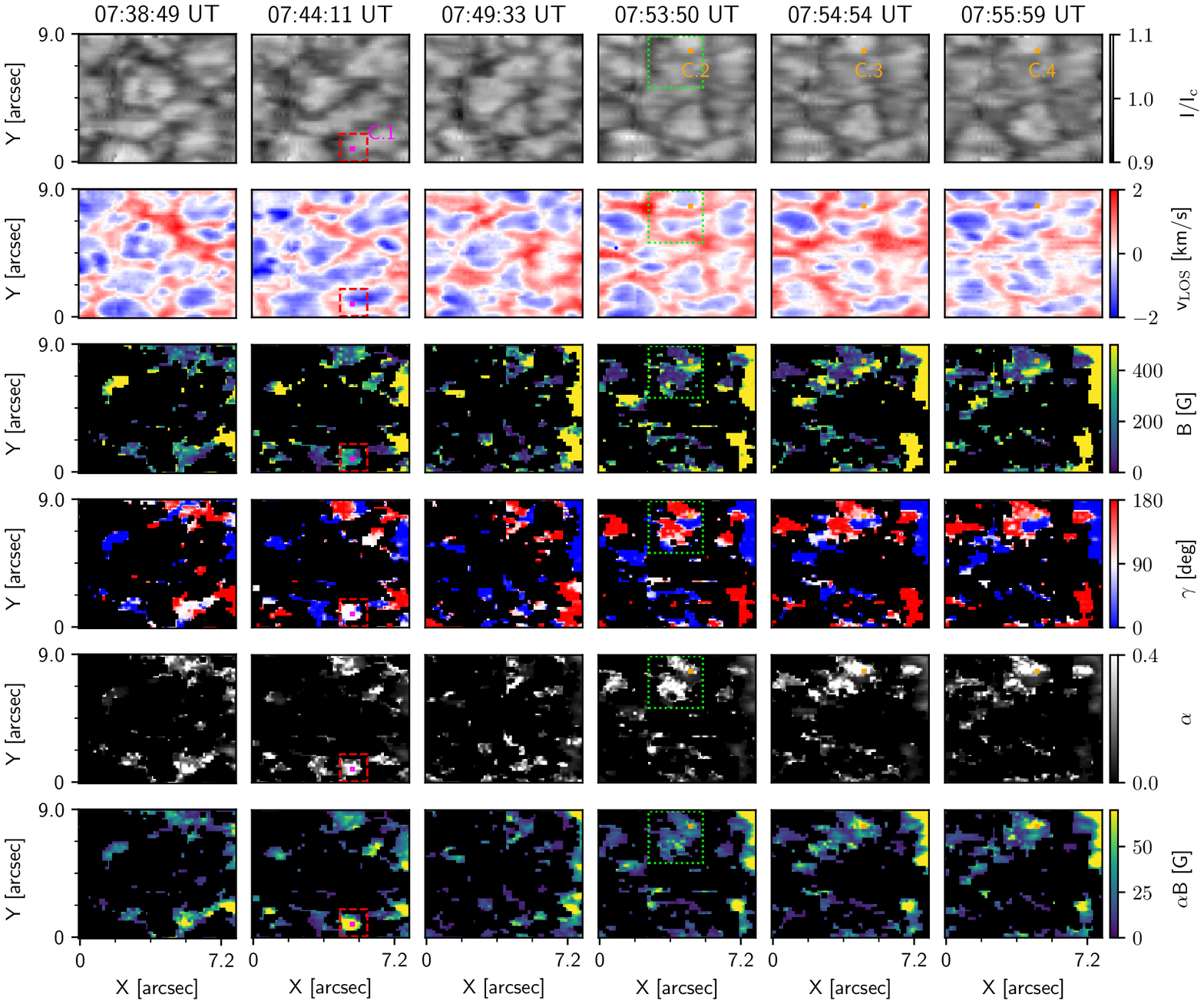}
\caption{ROI C: As in Figure \ref{fig:ROIA}, but for a region with several magnetic features visible in the time-series of the $6$ May scan, including a short-lived linear polarization feature (LPF), a complex weak magnetic structure with mixed polarities and transverse fields and a longitudinal kilo-Gauss magnetic element.}
    \label{fig:ROIC}
\end{figure*}

\begin{figure*}
    \centering
    \includegraphics[width=\textwidth]{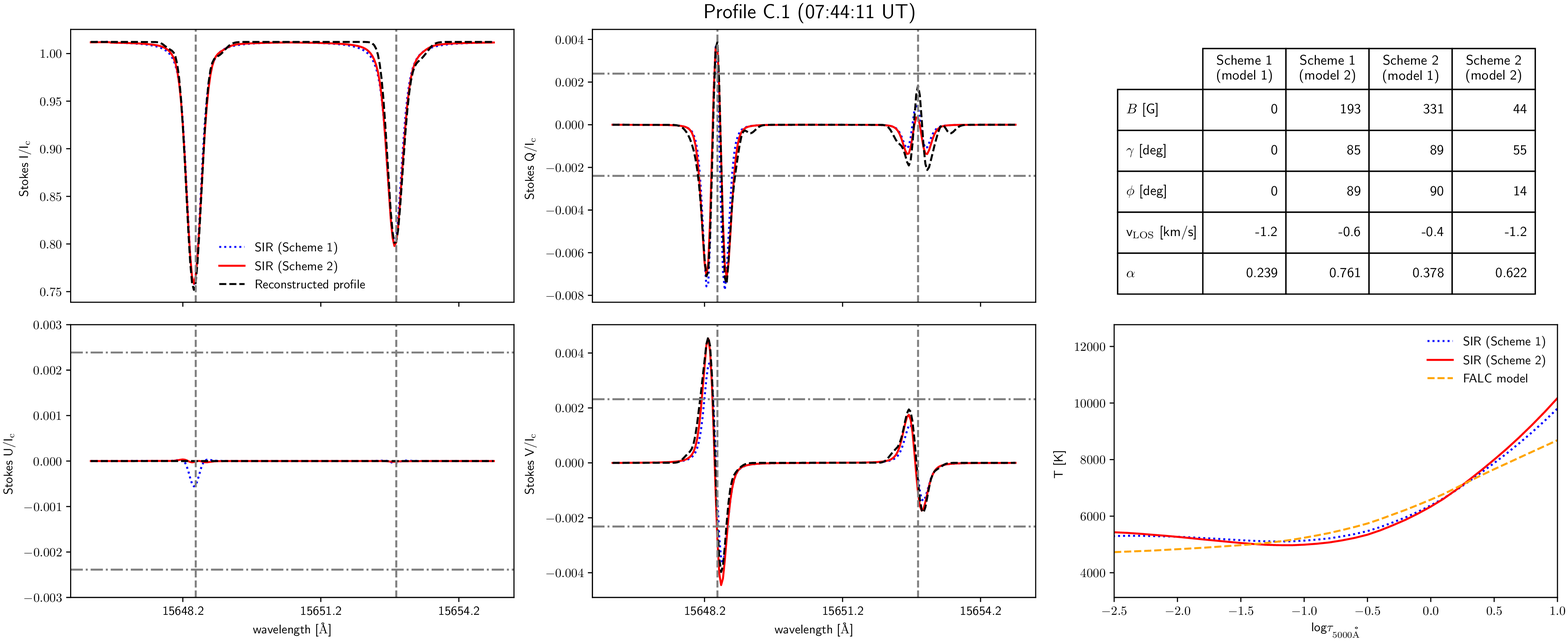}
    \caption{Profile C.1: As in Figure \ref{fig:ROIA_pix}, but for the pixel whose temporal and spatial location in the LPF present in ROI C is highlighted in Figure \ref{fig:ROIC}.}
    \label{fig:ROIC_pix}
\end{figure*}

Figure \ref{fig:ROIA_pix} shows the profiles labelled as A.1-3, which have all three polarized Stokes vectors satisfying the $\sigma_t$ level in each case, with the corresponding outputs from S1 and S2 inversions. In the first profile, A.1, the S1 inversion provides a good fit, revealing a highly inclined field ($\gamma = 82^\circ$) with a weak field strength ($B = 305$ G) such that $\alpha B = 60$ G. However, this scheme is unable to model the multiple lobes in the red-lobe of Stokes $V$. S2, on the other hand, is able to provide an improved fit and successfully models the multi-lobed nature of the Stokes $V$ profile. The SIR code has achieved this by superimposing two fields; one weaker ($B = 155$ G) highly inclined ($\gamma = 86^\circ$) field (with higher $\alpha$) and one stronger ($B = 295$ G) highly vertical ($\gamma = 18^\circ$) field (with lower $\alpha$). The thermodynamics of the two schemes are virtually identical, however, crucially, the two magnetic models in S2 have opposing $v_{\mathrm{LOS}}$, which is necessary to produce multiple lobes in Stokes $V$. Nevertheless, the resultant $v_{\mathrm{LOS}}$ is the similar in both schemes ($v_{\mathrm{LOS}} = 1.2-1.3$ km/s). While S2 is able to fit the multiple lobes, whether the red lobe can be trusted as a real signal or not is a different question; the red lobe does not reach the $\sigma_t$ threshold. In the next frame, profile A.2, we see that the highly inclined field has changed polarity. The Stokes $V$ profile is very weak, and barely reaches the $\sigma_t$ threshold. Interestingly, the asymmetry in the red and blue lobes of Stokes $Q$ and $U$ has become more pronounced, and the linear polarization is of a much higher amplitude. S2 achieves a moderately good fit to this asymmetry with an adjusted azimuth and significantly increased field strength in model 2. The kinematics of the former and latter frames are much the same. In the next frame, profile A.3, the Stokes $V$ profile is much stronger and two-lobed, such that S1 provides a reasonably good fit, and has changed polarity compared to A.1. On the other hand, S1 is unable to accurately fit the linear polarization, while S2 is able to provide a very good fit as it is able to adjust the velocities of its two highly inclined models accordingly. 

Figure \ref{fig:ROIB} shows a selection of frames from ROI B, observed in the $5$ May scan. This is a short-lived LPF that appears between and bridges two patches of circular polarization of the same polarity (enclosed by the dashed, cyan square in Figure \ref{fig:ROIB}). This LPF is located at the granule-IGL boundary of a very small granule, as is easiest to observe in the velocity maps. As in Figure \ref{fig:ROIA} for ROI A, Figure \ref{fig:ROIB} also shows the evolution of the magnetic and kinematic properties for the single pixel whose location is marked and labelled as B.1; this pixel is located at the centre of the LPF in the frame where the $\alpha B$ reaches its peak, and is shown in Figure \ref{fig:ROIB_pix}. The $\alpha B$ for this LPF rises to a peak of $133$ G at 07:56:29 UT before decaying, resulting in a lifetime of $6-7$ minutes. Even at its greatest extent, the LPF is never greater than $1''$ in diameter. During its lifetime, the kinematics of the feature are stable, but after the LPF decays the pixel experiences a strong down-flow; at that point, the widening of the IGL next to where the LPF was located is notable. From Figure \ref{fig:ROIB_pix}, it is clear that the amplitude of the linear polarization signals are very high, but the Stokes $V$ profile is weaker. This pixel, in this frame, is the only one during the lifetime of this LPF where the Stokes $V$ profile reaches the $\sigma_t$ noise threshold.  S1 is able to fit the linear polarization well with a highly inclined ($\gamma = 88^\circ$) weak field ($B = 283$ G) with relatively large $\alpha$ ($\alpha = 0.471$), but is not able to fit the multi-lobed Stokes $V$ profile. S2, on the other hand, is able to fit the small asymmetries between the amplitudes of the blue and red lobes of Stokes $Q$ and $U$ whilst providing a much improved fit to Stokes $V$ by combining two highly inclined fields ($\gamma = 84^\circ$ and $\gamma = 103^\circ$). If this Stokes $V$ profile is real, the question naturally follows whether the opposite polarities were also present in previous frames, but cancelled each other out such that the resultant Stokes $V$ amplitude did not reach the $\sigma_t$ noise threshold. It is likely that binning around this pixel would result in loss of information due to Zeeman cancelling, however it would also result in a lower noise level and thus allow us to determine whether there is indeed a Stokes $V$ signal in this LPF. Figure \ref{fig:ROIB_pix} also shows the profile (B.2) binned in the magenta square in Figure \ref{fig:ROIB}. The noise level is reduced, as one would expect, by $\sqrt{N_b}$, where $N_b = 4$ is the number of binned pixels. The atmospheric parameters retrieved by the S2 inversion for the binned and unbinned profiles are not significantly different, and the linear polarization profiles are very similar (albeit with slightly reduced amplitudes). Stokes $V$, however, is different; in the unbinned profile (B.1), we see only one lobe of Stokes $V$ above the $\sigma_t$ noise threshold, but two lobes in the binned profile (B.2) reach or exceed the $\sigma_t$ noise threshold, not just in the $g_{\mathrm{eff}}=3$ line but also the $g_{\mathrm{eff}}=1.5$ line. S1 provided a good fit to Stokes $I$, $Q$, and $U$ in both the unbinned and binned cases, but failed to fit Stokes $V$ in either case. Indeed, the synthetic Stokes $V$ profile is even of different polarity in the binned and unbinned profiles. 

Figure \ref{fig:ROIC} presents a selection of frames from ROI C, our final ROI, observed in the right-hand side of the full FOV of the scan observed on the $6$ May. At the beginning of the scan the magnetic structures are highly fragmented throughout, with no clear features. By 07:38:49 UT we begin to clearly see the emergence of an LPF at the bottom of the FOV, situated across a granule that has expanded in size (enclosed by the dashed, red square in Figure \ref{fig:ROIC}). Before these frames, there are some traces of very weak linear polarization in the immediate area, but it takes a few minutes for this LPF to become fully established. By 07:43:07 UT, this granule has become more elongated in shape and the LPF migrates more clearly to the granule-IGL boundary. In this frame the spatial extent of the LPF is around $1.5''$ in diameter. It is notable that at 07:38:49 UT the LPF seems to be attached to the strong vertical field to its right, but by 07:44:11 UT it has seemingly fully detached. Even after the LPF disappears by 07:49:33, this strong vertical field element persists throughout the scan. The lifetime of this LPF is much longer than the LPF observed in ROI B, perhaps even as high as $16$ minutes, but selection of the exact moment of appearance in particular is subjective. Figure \ref{fig:ROIC_pix} shows an example profile, labelled and marked as C.1 in Figure \ref{fig:ROIC}, from this LPF. In the $g_{\mathrm{eff}}=3$ line, all lobes of Stokes $Q$ and $V$ are detected above the $\sigma_t$ noise threshold, but Stokes $U$ is not, and, thus, Stokes $U$ has been set to zero. One may nevertheless observe that SIR has inserted a very small magnitude Stokes $U$ profile in its synthetic fit; setting the vector to zero has minimized the influence of this behaviour on the result. We attempted to bin spatially around this pixel to lower the noise threshold but the Stokes $U$ profile still did not reach the threshold. From the polarization one can immediately infer that the magnetic field is clearly highly inclined but with a significant vertical component, and this is reflected in the inversion results. The weak field ($B = 193$ G) with large $\alpha$ ($\alpha = 0.761$) from S1 is notable, resulting in a large magnetic flux density ($\alpha B = 147$ G) higher than the value determined for ROI B. S2 does not improve the fit but achieves very similar synthetic profiles by combining a highly inclined ($\gamma = 89^\circ$) weak ($B = 331$ G) field with a weaker ($B = 44$ G) and intermediately inclined ($\gamma = 55^\circ$) field. The thermodynamics of both schemes are in close agreement, and the field is unambiguously in an up-flow.

\begin{figure*}
    \centering
    \includegraphics[width=\textwidth]{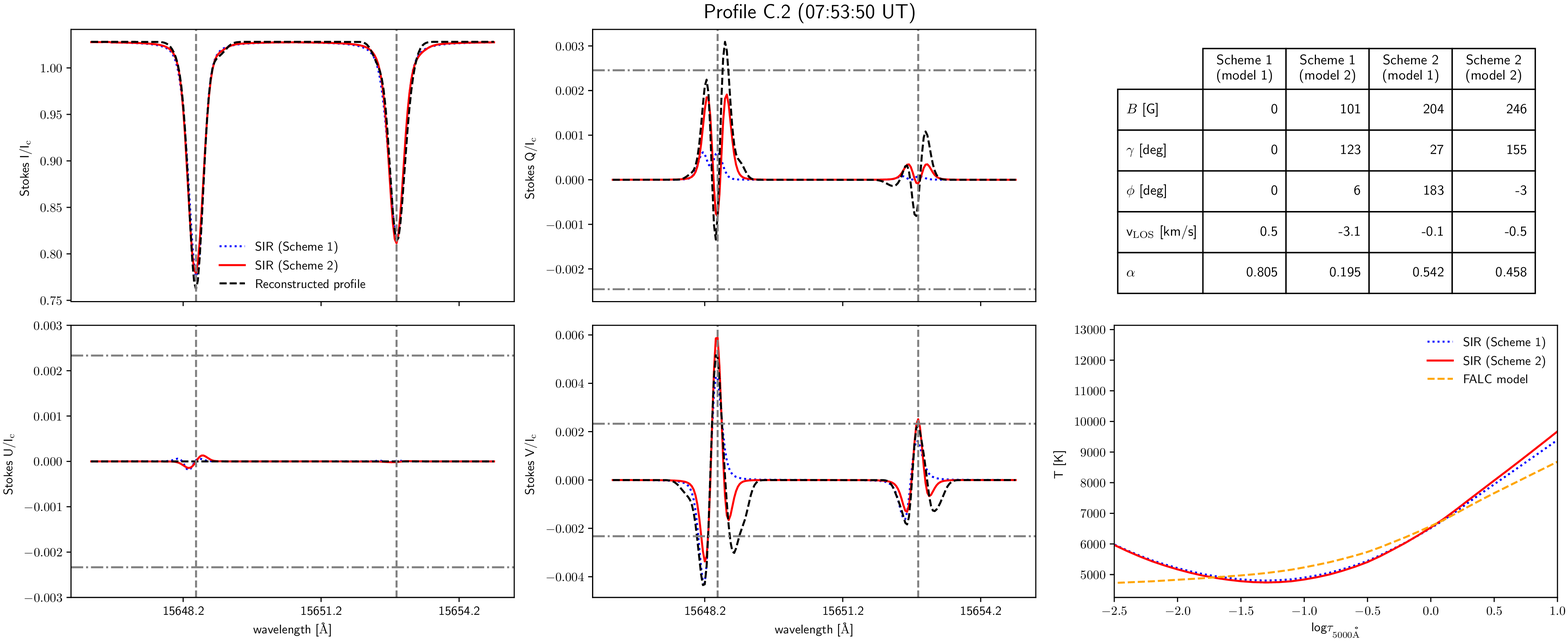}\vspace{0.1cm}
    \includegraphics[width=\textwidth]{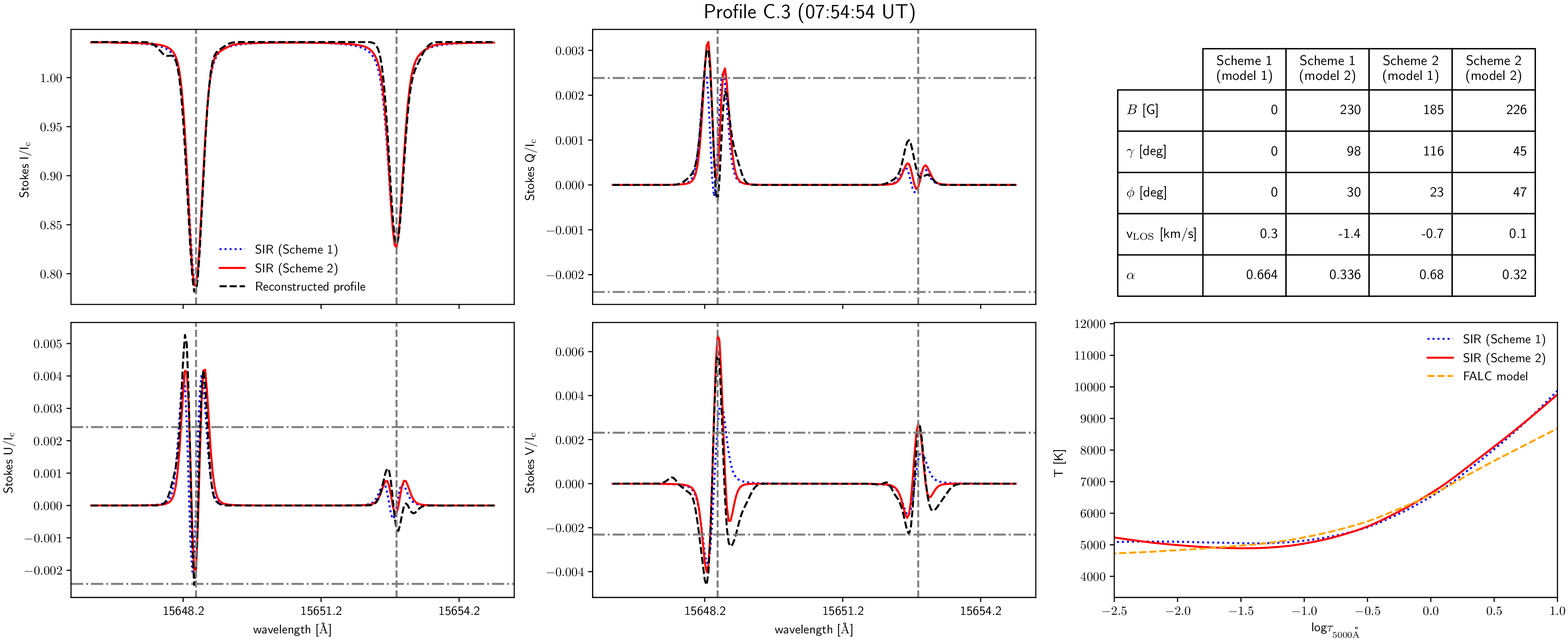}\vspace{0.1cm}
    \includegraphics[width=\textwidth]{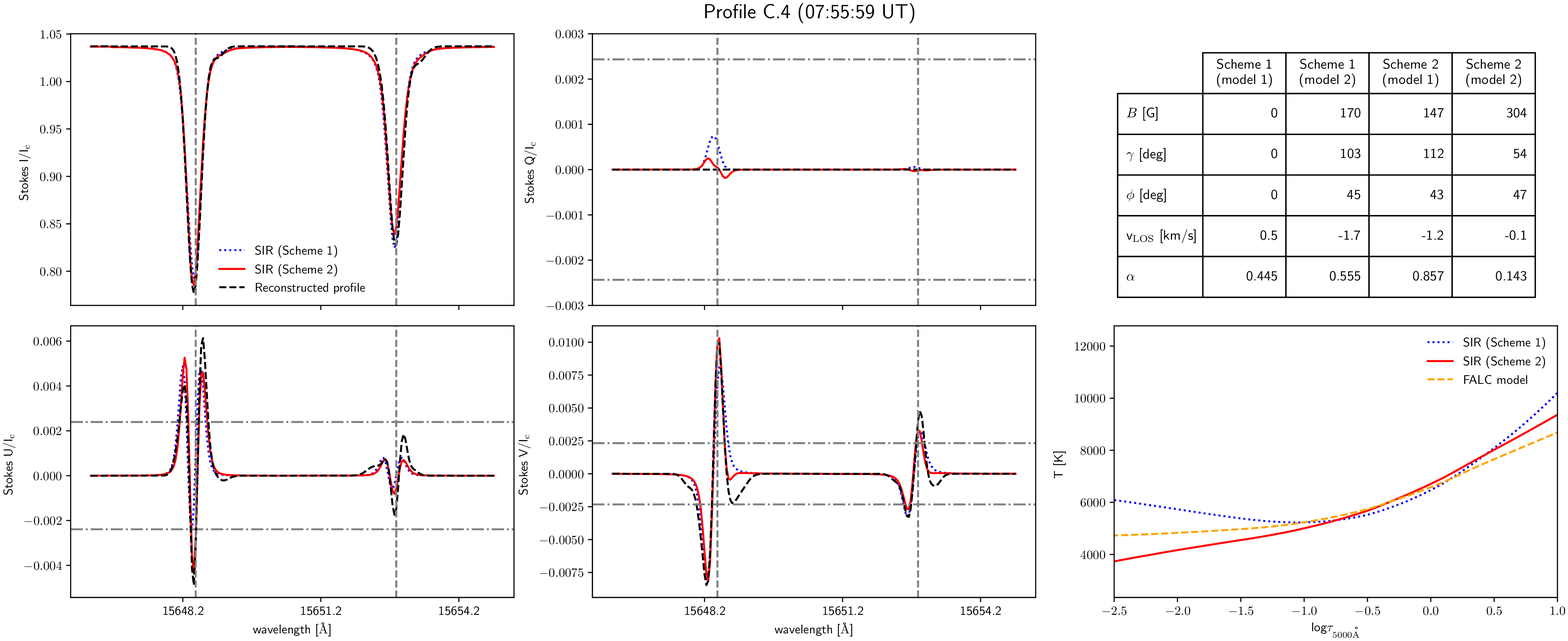}
    \caption{Profiles C.2-C.4: As in Figure \ref{fig:ROIA_pix}, but for the pixels whose temporal and spatial location in ROI C is highlighted in Figure \ref{fig:ROIC}.}
    \label{fig:ROIC_pix2}
\end{figure*}

\begin{figure*}
    \centering
    \includegraphics[width=\textwidth]{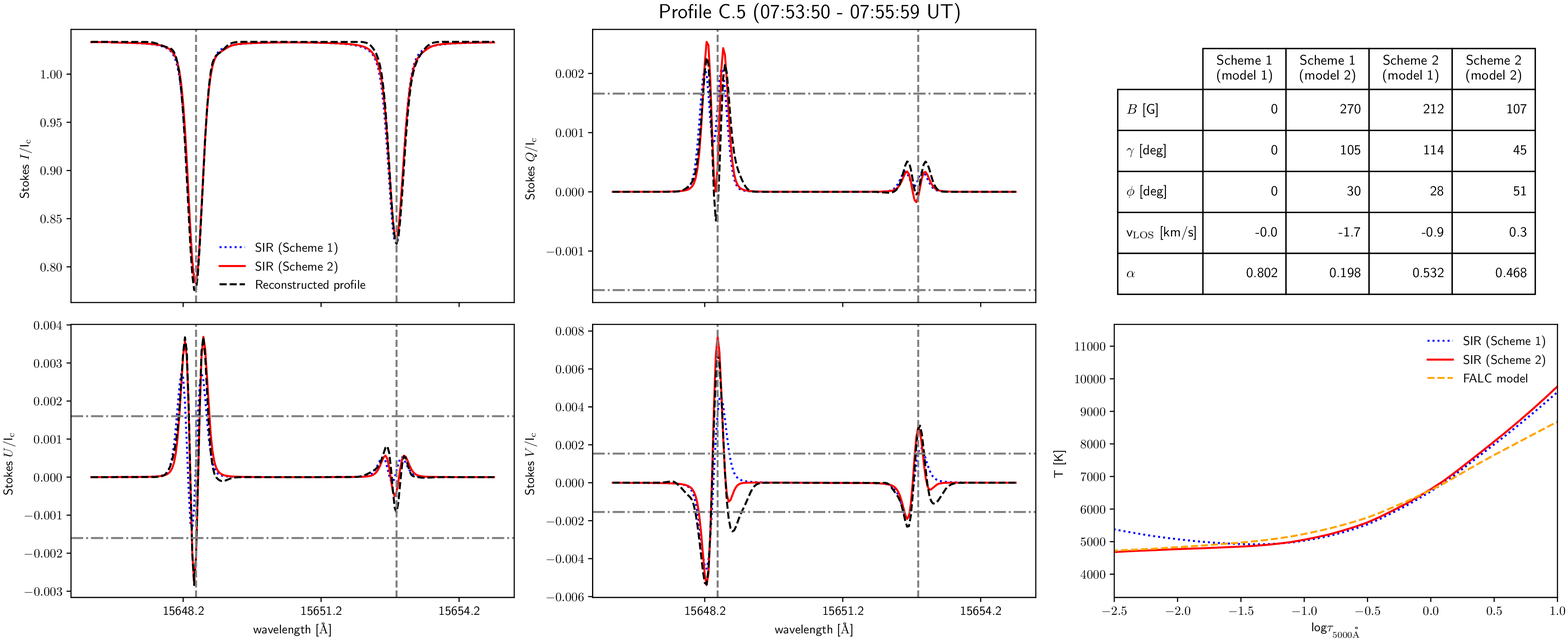}
    \caption{Profile C.5: As in Figure \ref{fig:ROIA_pix}, but for the pixels C.2-C.4, whose temporal and spatial location in ROI C is highlighted in Figure \ref{fig:ROIC}, binned temporally to produce the combined profile.}
    \label{fig:ROIC_pix3}
\end{figure*}

Located in the upper centre of ROI C is a magnetic structure visible as a complex mix of opposite polarity circular polarization separated by linear polarization (enclosed by the dotted, green square in Figure \ref{fig:ROIC}). In the 07:54:54 UT frame this structure is observed at its most continuous. The linear polarization is located at the granule-IGL boundary. Figure \ref{fig:ROIC_pix2} shows three example profiles (C.2-4) from a pixel whose location in this structure is marked in Figure \ref{fig:ROIC}, as observed at 07:53:50, 07:54:54 and 07:55:59 UT, and which has at least two polarized Stokes vectors above a $\sigma_t$ level in each case, with the corresponding outputs from S1 and S2 inversions. Profile C.3 has all three polarized Stokes vectors above the $\sigma_t$ threshold, but Stokes $U$ and $Q$ in C.2 and C.4, respectively, do not meet the threshold. In the case of C.2, while one lobe of Stokes $Q$ does reach the threshold, this profile is very weak and as such the lobes of the synthetic profiles do not reach the threshold. This is not the case for C.3 and C.4, however, such that we can have confidence that the linear polarization signals are real. The three-lobed Stokes $V$ in C.2-4, on the other hand, bears resemblance to A.2 from ROI A. S2 makes an attempt at fitting all three Stokes $V$ lobes in C.2 and C.3, however it is not completely successful. According to S1, the $\alpha B$ values of A.2 and C.3 were returned as $85$ G and $77$ G, respectively. The difference, however, is that in the case of C.3 all Stokes $V$ lobes are equal to or above $\sigma_t$, and we can therefore have more confidence that there could be mixed-polarities within this resolution element. We manually checked the profiles above and below C.3 to confirm the presence of Stokes $V$ signals with opposite polarities and magnitudes well in excess of the $\sigma_t$ noise threshold. In ROI B, we binned spatially to reduce the noise level, and were able to do so as the LPF had a degree of spatial continuity. In this case, opposite polarity Stokes $V$ profiles appear in close proximity such that the polarity of the Stokes $V$ profile would be fundamentally altered by spatial binning. However, we do have a sense of continuity in polarity between frames, as is evident from profiles C.2-4. Figure \ref{fig:ROIC_pix3} shows the temporally binned profile. All polarized Stokes lobes are above the reduced $\sigma_t$ noise level, with the exception of the Stokes $Q$ $\pi$ lobe. As in C.2-4, S1 provides a reasonable fit but the S2 synthetic profiles are an improvement. The value returned from S1 for $\alpha B$ is somewhat lower at $53$ G.

\section{Discussion and conclusions}\label{section:conclusions}

We have observed in the deep photosphere, for the first time with time-series imaging at very high-resolution in the near infrared, the temporal evolution of the small-scale internetwork magnetic field. After noise reduction and removal of instrumental artefacts, we find that $35\%$ of our FOV has either LP or CP above a $\sigma_t$ level. We compared the level 1 results to GRIS/GREGOR and IMaX/SUNRISE datasets and found similar statistics in LP and CP compared to the former dataset. Given that the visible line observed by IMaX/SUNRISE and the near infrared line observed by GRIS-IFU/GREGOR have the same $g_{\mathrm{eff}}$, we can use equation \ref{eqn:VtoB} to provide a very simple argument to underpin our assertion that the longer wavelength of the near infrared Fe lines is the most significant reason for the increase in polarization recorded by the GRIS-IFU over those obtained by IMaX (see Table \ref{table:1} and \ref{table:reconstructedstats}). According to the Zeeman ratio of these lines, all else except $\lambda_0$ being equal in equation \ref{eqn:VtoB}, one would expect three times more LP and CP to be observed in the near infrared. We refer the reader to the full explanation in \cite{marian2008b}. The LP recorded by the GRIS-IFU, after reconstruction, is in fact exactly as would be expected, while the CP is marginally lower. Although both the GRIS-IFU and IMaX instruments observed with essentially the same $1\sigma$ noise level, this small discrepancy is unsurprising given that the GRIS-IFU observed at lower resolution than IMaX, which has a detrimental impact on the measurement of CP in particular due to the mixing of polarities within the resolution element. It is important to note that the improvements to the adaptive optics and de-rotator systems at GREGOR since 2016 are factors acting in the GRIS-IFU's favour relative to GRIS \citep{kleint2020}. When comparing the results of either GRIS or the GRIS-IFU to IMaX, it would be remiss not to mention the impact of the atmosphere. For IMaX, the polarization ratios are remarkably stable throughout the time-series (see Figure \ref{fig:temporal}), perhaps due to the reduction in atmospheric turbulence. However, it is important to note the FOV of IMaX is far larger than the GRIS-IFU (nearly $14$ times bigger by area), which may therefore result in more temporally stable statistics. In addition, it is important to note the difference in formation heights between the visible and near infrared lines, with the former formed higher in the photosphere than the latter. The difference in stratification of the response of the Stokes vectors to $B$ can have important diagnostic consequences \citep{khomenko2007} and could influence the expected relative levels of LP and CP measured, especially in the context of an atmosphere whose density is understood to decrease as height increases.

We applied LTE inversions using SIR to the dataset, and statistically considered the magnetic properties. It is arguably challenging for the GRIS-IFU to provide statistics that are truly representative of the quiet internetwork due to a restricted FOV, however we argue that these statistics may be provided through observing temporally as the target evolves, which slit-spectropolarimeters like GRIS cannot do at high cadence except in sit-and-stare mode, and by taking multiple scans as reported here. The smaller peak in the $B$ distributions of both scans in Figure \ref{fig:inv_parameters} at {\raise.17ex\hbox{$\scriptstyle\mathtt{\sim}$}} $300$ G is similar to the {\raise.17ex\hbox{$\scriptstyle\mathtt{\sim}$}} $250$ G reported by \cite{marian} from inversions with a similar scheme. The value of the peak of the weak field population is significantly lower than the {\raise.17ex\hbox{$\scriptstyle\mathtt{\sim}$}} $450$ G peak reported by \cite{khomenko2003}, but their result was determined directly from the SFA, from splitting of the lobes of the Stokes vectors, which is an upper limit. The small $\alpha$ values returned are consistent with \cite{marian}, as one would expect for observations taken of similar targets with the same telescope. The ratio of the median of $B_\perp$ to the median of $B_\parallel$ is lower at $1.8-2.1$ than the ratio of averages recorded by \cite{suarez2012} for SP/Hinode, determined to be $3$ with the $6302$ $\AA$ Fe I line pair. It must be kept in mind that the median for $B_\perp$ is calculated across a much smaller number of profiles than the median for $B_\parallel$. One may propose that the lower ratio recorded in the near infrared relative to the visible lines could be understood under a simple qualitative understanding of the solar photosphere; it is expected that the distribution of $\gamma$ for magnetic flux elements must become more horizontal with height as ambient gas pressure decreases (density decreases).

As elucidated by \cite{borrero2011}, choosing a value for $\sigma_t$ that is too low can quickly result in an artificially inclined distribution of $\gamma$ values being returned from inversions due to noise contamination, even when synthesizing from magnetohydrodynamic
(MHD) outputs that contain only vertical fields. An indication that the $\sigma_t$ level is too low is the presence of a peak in the distribution of $\gamma$ at $90^\circ$, whereas with higher $\sigma_t$ levels this peak diminishes leaving two peaks either side of $\gamma = 90^\circ$. Following the advice of the authors, we therefore chose a conservative value for $\sigma_t$ for the reconstructed data that selects the same pixels as is selected by a $4-4.5\sigma$ level in the original level 1 data. In earlier iterations of our inversions, we retrieved $\gamma$ distributions with peaks at $\gamma = 30^\circ$ and $\gamma = 150^\circ$, and a very deep trough at $\gamma = 90^\circ$ (i.e. almost a complete absence of the spurious peak), suggesting that our $\sigma_t$ level was sufficiently stringent. This behaviour arises due to the tendency of inversion codes to insert spurious Stokes vectors with amplitudes lower or equal to the noise level. We therefore went further in our attempts to minimize this behaviour, and hence its influence on our statistical results, by setting individual polarized Stokes vectors to zero at all wavelengths if the $\sigma_t$ threshold was not satisfied. Our final $\gamma$ distribution, shown in Figure \ref{fig:inv_parameters}, has three distinct peaks at $\gamma = 0, 90, 180^\circ$. From Table \ref{table:reconstructedstats}, we can see only $3.8-4.6\%$ of pixels contained both LP and CP signals, and thus most polarized pixels contain either only Stokes $Q$ and $U$ (and thus highly inclined fields), or only Stokes $V$ (and thus highly vertical fields). The only way, therefore, to return a distribution for $\gamma$ that shows most pixels having both a strong vertical and horizontal component is if the inversion is contaminated by noise. Indeed, from the classification of inclinations returned by the inversion shown in Table \ref{table:inclin05}, we deduce that most of the fields measured at this spatial resolution and polarimetric sensitivity are highly vertical, although there is nevertheless evidence of a significant minority ($20-25\%$) of polarized profiles with a horizontal component to the field.

Finally, we defined, curated, and presented three ROIs and tracked the temporal evolution of small-scale magnetic features present in each. The GRIS-IFU is uniquely suited to this purpose. We typically find weak transverse fields in LPFs and complex, multi-polar loop-like structures with transverse fields. These magnetic structures are some of the smallest-scale loops ever observed in the quiet Sun. In ROI A we have clear loop-like structures with transverse fields separating opposite polarity longitudinal fields along the PIL. A key question therefore is whether there is enough observational evidence to determine whether these loops may be classified as U-shaped or $\Omega$-shaped. Theoretically the LOS kinematics of the transverse fields can provide a diagnostic for distinguishing the two cases, but as \cite{solanki2019} remark, it is difficult to disentangle changes to $v_{\mathrm{LOS}}$ due to cancellation processes from changes induced primarily by convective forces, due to the motion of flux that tends to emerge from granules before migrating to IGLs, where they tend to submerge. Therefore, although we clearly observe up-flows in profiles with transverse fields in the PIL (see Figure \ref{fig:ROIA_pix}), we cannot be sure this indicates a U-loop configuration as the PIL is located in a granule. Nevertheless, it is notable that in the last frame in Figure \ref{fig:ROIA_pix}, the $\alpha B$ has significantly diminished. In profile A.3, there is a clear incompatibility between the Doppler shift of Stokes $V$ versus that observed in Stokes $Q$ and $U$, which could indicate that $\alpha B_\perp$ is located at a different height than $\alpha B_\parallel$, or that we are not resolving the magnetic structures in the PIL. 

The LPF in ROI B serves as a good example for stressing the importance in simultaneously maximizing the spatial resolution, cadence and S/N of quiet Sun observations to prevent the mixing of signals within resolution elements. This magnetic structure is clearly organized on a scale smaller than we are capable of resolving, and thus are at risk of over-interpreting the Stokes profiles. We are therefore limited to broad interpretations; we can say with some certainty that the field is highly inclined, but not exclusively transverse, and its $B$ is weak. Comparing the first frame of Figure \ref{fig:ROIB} to the last, we initially observe relatively low $\alpha B$ spread across a larger area, which is followed by the appearance of the high flux density LPF. By the end of the time-series we are left with essentially exclusively $\alpha B_\parallel$ that is much higher and concentrated in a smaller area than at the beginning. It is plausible either that $\alpha B_\perp$ is conserved by submergence, although we only observe a down-flow after the LPF has disappeared, or conserved by the increase in $\alpha B_\parallel$.

In ROI C we observe another complex loop-like structure as well as another LPF, as in ROI A and B, respectively. The detection of a much stronger kilo-Gauss magnetic structure in the upper right of the FOV in Figure \ref{fig:ROIC} could have important consequences. This magnetic element first appears at 07:43:07 UT, forming along an IGL, and its $\alpha B$ increases in subsequent frames. The question then naturally arises whether the weak magnetic structures observed, not just in the $6$ May scan but also the $5$ May scan, are the decaying products of much stronger magnetic elements, that could be located just outside the FOV. Conversely, it could also be the case that the weak internetwork magnetic structures merge with the stronger network elements, as \cite{Gosic2014} observed commonly occuring in a statistical study of Hinode magnetograms. Along the PIL of the loop-like structure we observe a Stokes $V$ profile (C.5, see Figure \ref{fig:ROIC_pix3}), with three lobes above the $\sigma_t$ level after temporal binning, a signature of mixed polarities. We also observe this in ROI A, although not above $\sigma_t$ (A.1, see Figure \ref{fig:ROIA_pix}). Therefore, we have found linear polarization located between opposite polarity vertical fields in both ROI A and C, and it is certainly conceivable that this may be characteristic of how the deep photospheric magnetic field is organized at these scales and resolution. Mixed polarity Stokes $V$ profiles have also been observed in the near infrared in previous studies (\citet{khomenko2003,marian,kiess}). Multi-lobed Stokes $V$ profiles have been found in highly blue- or red-shifted Stokes profiles, indicative of supersonic magnetic flows at possible sites of reconnection \citep{Borrero2013}. However, the $v_{\mathrm{LOS}}$ values derived from our inversions do not support such a conclusion. As ever, we cannot, however, rule out the presence of noise. Residual cross-talk between the polarized vectors induced by environmental polarization is unlikely to provide an explanation as we do not observe the same spectral features in other pixels within the same IFU tile.

The clear advantage of IFUs is clear; the instrument enabled us to observe the dynamics of these photospheric structures and interrogate them on small scales in an unprecedented way. We pushed the detection capabilities of the GRIS-IFU to the limit by additionally spatially and temporally binning in select cases to reveal the presence of unresolved mixed polarity fields within the resolution element. In future, multi-wavelength and multi-instrument observations will be desirable as these will be required to constrain, at potential cancellation sites in the quiet sun, whether $\alpha B$ is conserved by emergence, submergence or cancellation and, if the latter, whether any signature of cancellation (e.g. Ohmic heating) can be detected higher in the atmosphere. The design of the observational sequence reported in this study and its success in producing meaningful results should help guide future quiet Sun observations with IFUs. We therefore conclude that the next generation of high-resolution ground-based solar telescopes, the Daniel K. Inouye Solar Telescope (DKIST) \citep{DKIST} and the European Solar Telescope (EST) in particular, will be required to unveil further small-scale magnetic fields, particularly those with transverse components, in the internetwork photosphere hidden to the Zeeman effect at current resolutions. 

\begin{acknowledgements}
      The authors would like to thank Carlos Dominguez-Tagle whose pioneering work with the excellent GRIS-IFU made these observations possible. We express our appreciation also to all the engineering, operating, and technical staff at GREGOR for their assistance during the observing campaign, including observing assistants Karin Gerber and Oliver Wiloth, in addition to Anjali John Kaithakkal for sharing her expertise at the beginning of the campaign. Gratitude is extended to Elena Khomenko, Marian Martínez González and Juan Manuel Borrero for their advice and numerous insightful discussions. RJC would also like to thank Valentin Mart\'inez Pillet for making the IMaX data available. RJC thanks Robert Ryans for IT support and assistance in utilizing QUB's high performance computing (HPC) facilities. Although the simultaneously recorded data was not published, we thank Carsten Denker and Christoph Kuckein for assistance with operating the High-resolution Fast Imager (HiFI) instrument and associated data reduction. This research has received financial support from the European Union’s Horizon $2020$ research and innovation program under grant agreement No. $824135$ (SOLARNET). RJC acknowledges support from the Northern Ireland Department for the Economy (DfE) for the award of a PhD studentship. MM, CJN and AR acknowledge support from the Science and Technology Facilities Council (STFC) under grant No. ST$/$P000304$/$1 $\&$ ST$/$T00021X$/$1. D.K. has received funding from the S\^{e}r Cymru II scheme, part-funded by the European Regional Development Fund through the Welsh Government and from the Georgian Shota Rustaveli National Science Foundation project FR17\_\ 32. The $1.5$-meter GREGOR solar telescope was built by a German consortium under the leadership of the Leibniz-Institute for Solar Physics (KIS) in Freiburg with the Leibniz Institute for Astrophysics Potsdam, the Institute for Astrophysics Göttingen, and the Max Planck Institute for Solar System Research in Göttingen as partners, and with contributions by the Instituto de Astrof\'isica de Canarias and the Astronomical Institute of the Academy of Sciences of the Czech Republic. Helioseismic and Magnetic Imager (HMI) magnetograms, courtesy of NASA/SDO and the AIA, EVE, and HMI science teams, were used during observations for target selection.
      \end{acknowledgements}

%
%

\bibliographystyle{aa.bst} 
\bibliography{refs} 


\begin{appendix}
\section{Response functions}\label{section:RFs}

\begin{figure*}
    \centering
    \includegraphics[width=\textwidth]{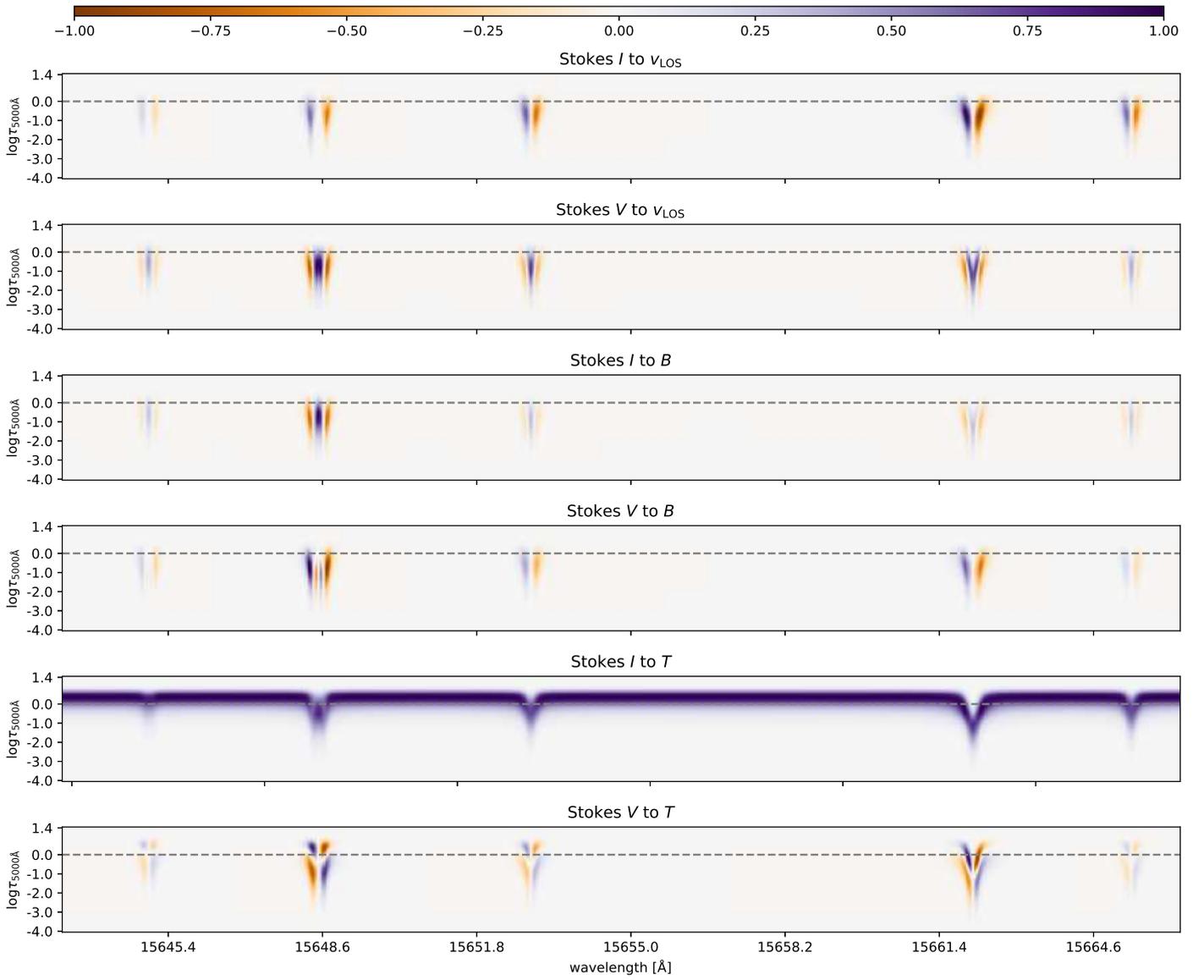}
    \caption{Response functions of Stokes $I$ and $V$ to $v_{\mathrm{LOS}}$, $B$ and $T$ in wavelength and optical depth, $\tau$. Each response function is normalized with respect to the maximum absolute value. The \textit{dashed} horizontal line emphasizes the location of log$\tau_{5000\mathrm{\AA}} = 0$. These response functions were computed by SIR, using profiles produced in synthesis mode from a FALC model atmosphere with $\gamma$ and $\phi$ fixed at $45^\circ$, $B$ at $300$ G and $v_{\mathrm{LOS}}$ at $0$ km/s.}
    \label{fig:RFs}
\end{figure*}

When determining an optimum inversion scheme it is an informative exercise to examine the response functions of the observed spectral lines \citep{RFs1996}. Response functions are a tool that allow us to understand the extent to which we can evaluate any quantity from an inversion and they can be analytically derived by SIR. We employ SIR in synthesis mode to produce synthetic Stokes profiles from the FALC model atmosphere, for all five spectral lines, fixing $\gamma$ and $\phi$ at $45^\circ$ and $B$ at $300$ G. Figure \ref{fig:RFs} shows the response functions of $B$, $T$ and $v_{\mathrm{LOS}}$ in optical depth and wavelength. We find that maximum response to $B$ is found at an optical depth of log$\tau_{5000\mathrm{\AA}} = -0.6$. The position of this maximum is invariant whether we integrate across all Stokes vectors, or Stokes $I$ and the polarized vectors separately. Further, this response is over a narrow range, with $75\%$ of the response to $B$ (and $\gamma$, although not shown) located between log$\tau_{5000\mathrm{\AA}} = 0.0$ and log$\tau_{5000\mathrm{\AA}} = -1.5$. In SIR, nodes must be evenly spaced throughout the atmosphere. With such narrow response functions, there is therefore little to gain by the introduction of gradients to $B$ (or $\gamma$). Further, as we are inverting unresolved magnetic structures and weak profiles, we elect to keep the number of free parameters minimized as we wish to prevent over-interpretation. From the response functions, we also determine, by incrementally increasing $B$, that the importance of Stokes $I$ relative to Stokes $Q$, $U$ or $V$ increases as $B$ increases, underlining the importance of both Stokes $I$ and the polarized vectors to field strength in a fully magnetized plasma. Evidently, the continuum of Stokes $I$ dominates the response to $T$, and the maximum response of Stokes $I$ to $T$ is found deep in the photosphere at an optical depth of log$\tau_{5000\mathrm{\AA}} = 0.4$. This spectral region is therefore a useful diagnostic for temperature in the deep photosphere \citep{borrero2017}. However, at wavelengths containing absorption lines the response of Stokes $I$ to T is stronger higher in the atmosphere and the response of Stokes $V$ to T is stronger lower in the atmosphere. There is some evidence that the lines are responsive to perturbations in $v_{\mathrm{LOS}}$ deep in the atmosphere \citep{ivan2019}, but we do not investigate this as it would require precise placement of nodes at given optical depths and SIR does not have the required functionality. Nevertheless it is clear that including all five lines can help constrain the inversion; for example, the response of Stokes $I$ to $v_{\mathrm{LOS}}$ is strongest in the $15662.02$ $\AA$ line, but the response of Stokes $V$ to $v_{\mathrm{LOS}}$ is strongest in the $15648.52$ $\AA$ line. We do not expect to strongly constrain the atmosphere beyond log$\tau_{5000\mathrm{\AA}} = -2.0$.

\section{Weak field approximation}\label{section:WFA}
The broadening produced by the Zeeman effect relative to the broadening due to thermal motions, encompassed by the parameter $\lambda_B/\lambda_d$, provides three regimes in which magnetic fields may be characterized: weak fields (i.e. $\lambda_B/\lambda_d \ll 1$), strong fields (i.e. $\lambda_B/\lambda_d \gg 1$) and a transitional intermediate case (i.e. $\lambda_B/\lambda_d \approx 1$). Further, it is necessary to distinguish each case by the degree to which they are spatially resolved, quantified by $\alpha$ ($\alpha=1$ when fully resolved). When the Zeeman splitting is negligible relative to the Doppler width of a line, when $\lambda_B/\lambda_d \ll 1$, and when $B$, $\gamma$, $\phi$ and $v_{\mathrm{LOS}}$ are invariant with height (i.e. no gradients) in the formation region of the line, we can say a given line is in the weak field regime when
\begin{equation}
    B < \frac{4\pi m}{g_{\text{eff}}\lambda_0 e}\sqrt{\frac{2kT}{M} + v^2_{mic}},
\end{equation}
where M is the mass of the species and $k$  is the Boltzmann constant. For the Fe line at $15648.52 \AA$ and assuming $v_{\text{mic}} = 1$ km s$^{-1}$, we estimate a weak field regime limit of around $270$ G for a typical photospheric temperature of $6500$ K. On the other hand, when the Zeeman splitting is larger than the Doppler width, when $\lambda_B/\lambda_d \gg 1$, through equation \ref{eqn:VtoB} we can measure $B$ directly from the separation of the $\sigma$-lobes of the polarized Stokes vectors. We refer to this as the strong field approximation (SFA). Series expansion of the absorption matrix allows the following expression for Stokes $V$ to be derived \citep{jefferies1989,PolSpectLines}:
\begin{equation}
    V(\lambda) \approx -\alpha \lambda_B \cos\gamma \frac{dI_0}{d\lambda},
\end{equation}
where $I_0$ is the emergent intensity if $B$ was zero. In the weak regime, the amplitude of the Stokes $V$ signal is proportional to the longitudinal field ($B_\parallel = B\cos\gamma$) when $\alpha=1$, or to the longitudinal magnetic flux density, $\alpha B_\parallel$, when $\alpha\neq1$,
\begin{equation}
    V(\lambda) \propto \frac{dI_0}{d\lambda} \times \alpha B_\parallel.
\end{equation}
Through equation \ref{eqn:VtoB}, we can therefore use the derivative of Stokes $I$ to estimate $\alpha B_\parallel$, or $B_\parallel$ under the assumption that $\alpha = 1$ or is known, assuming any spatially resolved magnetic features have the same $B$ and polarity. We refer to this as the weak field approximation (WFA). As discussed in \cite{solanki1993}, for spatially unresolved (i.e. $\alpha \neq 1)$ weak fields typically found in the quiet photospheric internetwork, the Zeeman effect may provide lower and upper limits on the field strengths using the WFA and SFA, respectively.  
\end{appendix}
\end{document}